\documentclass[a4paper]{amsart}
\usepackage{a4wide}
\usepackage[english]{babel}
\usepackage{amsmath,amsthm,amssymb}
\usepackage{graphicx}
\usepackage{subfigure}
\usepackage{enumitem}
\usepackage{tikz}
\usepackage{url}
\allowdisplaybreaks[3]

\newcommand{\sgn}{\text{ sign}}
\DeclareMathOperator{\Div}{div}

\begin{document}

\title[Opinion dynamics: inhomogeneous Boltzmann-type equations]{Opinion dynamics: inhomogeneous Boltzmann-type equations
  modelling opinion leadership and political segregation}
\author{Bertram D\"uring$^{1}$ and Marie-Therese Wolfram$^2$}
\address{$^1$ Department of Mathematics, University of Sussex,
Brighton, BN1 9QH, United Kingdom, Email: b.during@sussex.ac.uk\\
$^2$ Radon Institute for Computational and Applied Mathematics,
Austrian Academy of Sciences, 4040 Linz, Austria, Email: mt.wolfram@ricam.oeaw.ac.at}
%\subject{xxx}
%\corres{xx}
%\email{b.during@sussex.ac.uk, mt.wolfram@ricam.oeaw.ac.at}

\begin{abstract}
We propose and investigate different kinetic models for opinion
formation, when the opinion formation process 
depends on an additional independent variable, e.g.\
a leadership or a spatial variable.
More specifically, we consider:
(i) opinion dynamics under the effect of opinion leadership, where each
individual is characterised not only by its opinion, but also by
another independent variable which quantifies leadership qualities;
(ii) opinion dynamics modelling political segregation in the `The Big Sort', a phenomenon that US
citizens increasingly prefer to live in neighbourhoods with
politically like-minded individuals.
Based on microscopic opinion
consensus dynamics such models lead to inhomogeneous Boltzmann-type
equations for the opinion distribution. We derive macroscopic
Fokker-Planck-type equations in a quasi-invariant opinion limit and
present results of numerical experiments.
\end{abstract}

\maketitle

%%%%%%%%%%%%%%%%%%%%%%%%%%%%%%%%%%%%%%%%%%%%%%%%%%%
\section{Introduction}

The dynamics of opinion formation have been studied with growing
attention; in particular in the field of physics \cite{Gal05,DeAmWeFa02,SznSzn00},  in which a new research
field termed {\em sociophysics\/} (going back to the
pioneering work of Galam {\em et al.} \cite{GaGeSh82}) emerged. 
More recently, different kinetic models to describe opinion formation have 
been proposed
\cite{Tosc06,BerDel08,DuMaPiWo09,BoMoSa10,BorLor13a,BorLor13b,MotTad14}. 
Such models successfully use methods from statistical mechanics 
to describe the behaviour of a large number of
interacting individuals in a society.
This leads to generalisations of the classical Boltzmann
equation for gas dynamics.
Then the framework from classical kinetic theory for homogeneous gases is
adapted to the sociological setting by replacing molecules and their
velocities by individuals and their opinion. Instead of binary
collisions, one considers the process of compromise between two individuals.

The basic models typically assume a homogeneous society.
To model additional sociologic effects in real societies, e.g.\  the
influence of strong opinion leaders \cite{DuMaPiWo09}, one needs to consider {\em
  inhomogeneous\/} models. One solution, which arises naturally in certain
situations (see, e.g.\ \cite{Due10}), is to consider the time-evolution of distribution
functions of different, interacting species.
To some extent this can be seen as the analogue to the physical
problem of a mixture of gases, 
where the molecules of the different gases exchange momentum during collisions
\cite{BG}. This leads to systems of Boltzmann-like equations for the
opinion distribution functions $f_i=f_i(w,t),$ $i=1,\dots,n, $ of $n$
interacting species which are
of the form
\begin{equation}
\label{genB}
\frac{\partial}{\partial t}f_i(w,t)=\sum_{j=1}^n \frac1{\tau_{ij}} \mathcal{Q}_{ij}(f_i,f_j)(w).
\end{equation}
The Boltzmann-like collision operators $\mathcal{Q}_{ij}$ describe the change
in time of $f_i(w,t)$ due to binary interaction depending on a balance
between the gain and loss of individuals with opinion $w$. The
suitably chosen relaxation times $\tau_{ij}$ allow to control
the interaction frequencies. To model exchange of individuals (mass)
between different species, additional 
collision operators can be present on the right hand side of
\eqref{genB}, which are reminiscent of chemical reactions in the
physical situation.

Another alternative is to study
models where the distribution function depends on an
additional variable as e.g.\ in \cite{DuTo07}. This leads to {\em inhomogeneous
Boltzmann-type equations\/} for the distribution function $f=f(x,w,t)$ which
are of the following form
\begin{equation}
\label{eq:inhB0}
\frac{\partial}{\partial t}f+\Div_x(\Phi(x,w) f)=\frac1\tau \mathcal{Q}(f,f).
\end{equation}
Clearly, the choice of the field $\Phi=\Phi(x,w)$ which describes the
opinion flux plays a crucial role. It may not be easy
to determine a suitable field from the economic or sociologic
problem, in contrast to the physical situation where the law of
motion yields the right choice.

 In this paper, we give two examples of opinion formation problems,
 which can be modelled using {\em inhomogeneous
Boltzmann-type equations\/}.
One is concerned with opinion formation where the compromise process depends
on the interacting individuals' leadership abilities. The other
considers the so-called `Big Sort phenomena', the clustering of individuals with
similar political opinion observed in the USA.
Both problems lead to an inhomogeneous Boltzmann-type equation of the form
\eqref{eq:inhB0}.

% Toscanis approach and related\ldots
Our work is based on a homogeneous kinetic model for opinion formation introduced
by Toscani in \cite{Tosc06}. The idea of this kinetic model is to describe the
evolution of the distribution of opinion by means of {\em microscopic\/}
interactions among individuals in a society.
Opinion is represented as a continuous
variable $w\in\mathcal{I}$ with $\mathcal{I}=[-1,1]$, where $\pm 1$
represent extreme opinions. If concerning political opinions
$\mathcal{I}$ can be identified with the left-right political
spectrum. Toscani bases his model on two main aspects of opinion formation. The
first one is a {\em compromise process} \cite{HegKra02,DeAmWeFa02,Wei00}, in which individuals tend to
reach a compromise after exchange of opinions. The second one is {\em
  self-thinking,} where individuals change their opinion in a
diffusive way, possibly influenced by exogenous information sources
like the media. Based on both Toscani \cite{Tosc06} defines a kinetic model
in which opinion is exchanged between individuals through pairwise
interactions: when two individuals with pre-interaction opinion $v$ and $w$
meet, then their post-trade opinions $v^*$ and $w^*$ are given by
\begin{align}\label{e:toscani}
v^* = v - \gamma P(|v-w|)(v-w) + \tilde\eta D(v),\quad
w^* = w - \gamma P(|v-w|) (w-v) + \eta D(w). 
\end{align}
Herein, $\gamma \in(0, 1/2)$ is the constant {\em compromise
  parameter}. The quantities $\tilde\eta$ and $\eta$ are random variables
with the same distribution with mean zero and variance $\sigma^2$. They model {\em self-thinking\/} 
which each individual performs in a random diffusion fashion through an
exogenous, global access to information, e.g.\ through the press,
television or internet. 
The functions $P(\cdot)$ and $D(\cdot)$ model the local relevance of
compromise and self-thinking for a given opinion.
To ensure that post-interaction opinions remain in the interval
$\mathcal{I}$ additional assumptions need to be made on
the random variables and the functions $P(\cdot)$ and $D(\cdot)$, see
\cite{Tosc06} for details. 
In this setting, the time-evolution of the distribution of opinion 
among individuals in a simple, homogeneous society is governed by a
homogeneous Boltzmann-type equation of the form \eqref{genB}.
In a suitable scaling limit, a partial differential
equation of Fokker-Planck type is derived for the distribution of
opinion. Similar diffusion equations are also obtained
in \cite{SlaLav03} as a mean field limit the Sznajd model \cite{SznSzn00}.
Mathematically, the model in \cite{Tosc06} is related to works in the
kinetic theory of granular gases \cite{CeIlPu94}. In particular, the
non-local nature 
of the compromise process is analogous to the variable coefficient of
restitution in inelastic collisions \cite{Tosc00}. Similar models
are used in the modelling of wealth and income distributions which
show Pareto tails, cf.\ \cite{DuMaTo08} and the references therein.

The paper is organised as follows. We introduce two
new, inhomogeneous models for opinion formation.
In Section~\ref{sec:leadership} we consider
opinion formation dynamics which take into account the effect of
opinion leadership. Each individual is not only characterised by its
opinion but also by another independent variable, the assertiveness, which quantifies
its leadership potential. Starting from a microscopic model for the
opinion dynamics we arrive at an inhomogeneous Boltzmann-type
equation for the opinion distribution function. 
We show that, alternatively similar dynamics can be modelled
by a multi-dimensional kinetic opinion formation model.
In Section~\ref{sec:bigsort} we turn to the modelling of `The Big
Sort', the phenomenon that US citizens increasingly prefer to live
among others who share their political opinions. We propose a kinetic
model of opinion formation which takes into account these
preferences. Again, the time evolution of the opinion distribution is described by
an inhomogeneous Boltzmann-type equation. 
In Section~\ref{sec:fokkerplanck} we derive the corresponding macroscopic Fokker-Planck-type
limit equations for the inhomogeneous Boltzmann-type equations in a
quasi-invariant opinion limit.
Details on  the numerical solvers as well as results of numerical experiments are
presented in Section~\ref{sec:numerics}. Section \ref{sec:concl} concludes.

%%%%%%%%%%%%%%%%%%%%%%%%%%%%%%%%%%%%%%%%%%%%%%%%%%%

\section{Opinion formation and the influence of opinion leadership}
\label{sec:leadership}

The prevalent literature on opinion formation has 
focused on election processes, referendums or public opinion tendencies. 
With the exception of \cite{BerDel08,DuMaPiWo09}
less attention has been paid to the
important effect that opinion leaders have on the dissemination
of new ideas and the diffusion of beliefs in a society.
Opinion leadership is one of several sociological models trying to explain
formation of opinions in a society. Certain, typical personal characteristics are
supposed to characterise opinion leaders: high confidence, high
self-esteem, a strong need to be unique, and 
the ability to withstand criticism.
An opinion leader is socially active, highly connected and held in
high esteem by those accepting his or her opinion. 
Opinion leaders appears in such different areas as political
parties and movements, advertisement of commercial products and
dissemination of new technologies.

In the opinion formation model in \cite{DuMaPiWo09} the society is built of two groups, one group of
opinion leaders, and one of ordinary people, so-called
followers. In this model, individuals from the same group can influence each others
opinions, but opinion leaders are assertive and, although able to
influence followers, are unmoved by the followers' opinions. In this model a leader always
remains a leader, a follower always a follower. Hence it is not possible to describe the emergence
or deline of leaders.

In the model proposed in this section we assume that the leadership qualities (like
assertiveness, self-confidence, ...) of each individual are characterised
through an additional independent variable, to which we
refer short as {\em assertiveness}.
In the sociological literature the term assertiveness describes a person's
tendency to actively defend, pursue and speak out 
for his or her own values preferences and goals. There has been a lot of research on
how assertiveness is connected to leadership, see for example \cite{AmeFly07}
and references therein.

\subsection{An inhomogeneous Boltzmann-type equation}
\label{sec:inhomogBoltzmann}

Our approach is based on Toscani's model for opinion formation \eqref{e:toscani}, but assumes that the compromise 
process is also influence by the assertiveness of the interacting individuals.
The assertiveness of an individual is represented by the continuous variable $x \in \mathcal{J}$ with $\mathcal{J} = [-1,1]$. 
The endpoints of the interval $\mathcal{J}$, i.e.\ $\pm 1$, represent a strongly or weakly assertive individual. A leader 
would correspond to a strongly assertive individual.

The interaction for two individuals with opinion and assertiveness $(v,x)$ and $(w,y)$ reads as:
\begin{align}\label{e:exchange}
v^* = v - \gamma C(x,y,v,w)(v-w) + \tilde\eta D(v),\quad
w^* = w - \gamma C(x,y,v,w) (w-v) + \eta D(w). 
\end{align}
The first term on the right hand sides of \eqref{e:exchange} models the compromise process, 
the second the self-thinking process. 
The functions $C(\cdot)$ and $D(\cdot)$ model the
local relevance of compromise and self-thinking for a given opinion, respectively. 
The constant {\em compromise parameter\/} $\gamma \in(0,1/2)$ controls the `speed' of attraction of two
different opinions. The quantities $\tilde\eta,$ $\eta$ are random
variables with distribution $\Theta$ with variance $\sigma^2$ and
zero mean, assuming values on a set $\mathcal{B}\subset\mathbb{R}.$

Let us discuss the interaction described in \eqref{e:exchange} and its
ingredients in more detail.
In each such interaction, the pre-interaction opinion $v$ increases (gets closer to $w$) when
$v<w$ and decreases in the opposite situation;  the change of the
pre-interaction opinion $w$ happens in a similar way.
We assume that the compromise process function $C$ can be 
written in the following form:
\begin{align}
C(x,y,v,w) = R(x,x-y) P(\lvert v-w\rvert).
\end{align}
In \eqref{e:exchange} we will only allow interactions
that guarantee $v^*,w^*\in \mathcal{I}$. To this end, we assume
$$
0 \leq P(|v-w|) \leq 1,\quad 0 \leq R(x,x-y) \leq 1,\quad 0\leq D(v) \leq 1.
$$
Let us now discuss useful assumptions on the functions $P$, $R$ and
$D$.

As in \cite{Tosc06} we assume that the
ability to find a compromise is linked to the distance between
opinions. The higher this distance is, the lower the possibility to
find a compromise. Hence, the {\em localisation function\/} $P(\cdot)$
is assumed to be a decreasing function of its argument. Usual choices
are $P(|v-w|)=\mathbf{1}_{\{|v-w|\le c\}}$ for some constant $c>0$ and
smoothed variants thereof \cite{Tosc06}.

On the other hand, we assume the higher the assertiveness level, the
lower the tendency of an individual to change their opinion.  
Hence,  $R(\cdot)$ should be decreasing in its argument, too.
A possible choice for $R(\cdot)$ may be
\begin{align}\label{e:R}
 R(x,x-y)=R(x-y) = \frac{1}{2}-\frac{1}{2} \tanh\big(k(x-y)\big)
\end{align}
for some constant $k>0$.
This choice is motivated by the following considerations: Let $A$ and $B$ denote two individuals
with assertiveness and opinion $(x,v)$ and $(y,w)$, respectively. Then the particular choice of \eqref{e:R} corresponds
to the two cases: 
\begin{itemize}
\item If $x \approx 1$ and $y \approx -1$, i.e.\ a highly assertive individual $A$
meets a weakly assertive individual $B$: then $R(x-y) \approx 0$ (no influence of individual $B$
on $A$), but $R(y-x) \approx 1$, i.e.\ the leader $A$ persuades a weakly assertive individual $B$.
\item If both individuals have a similar assertiveness level and hence\ $\lvert
  x-y \rvert$ is small, there
will be some exchange of opinion, no matter how large (or small) this
assertiveness level is.
\end{itemize}
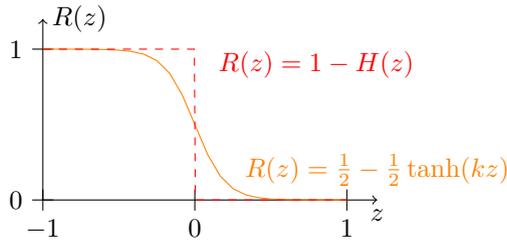
\begin{figure}\centering
\begin{tikzpicture}[domain=-1:1,scale=2]
  \draw[->] (-1,0) -- (1.2,0) node[below] {$z$};
    \draw[->] (-1,0) -- (-1,1.2) node[right] {$R(z)$};
\foreach \z/\ztext in {-1/-1,0/0,1/1}
    \draw[shift={(\z,0)}] (0pt,2pt) -- (0pt,-2pt) node[below]
    {$\ztext$};
\foreach \y/\ytext in {0/0,1/1}
\draw[shift={(-1,\y)}] (2pt,0pt) -- (-2pt,0pt) node[left] {$\ytext$};
    \draw[color=orange] plot[id=R] function{0.5-0.5*tanh(5*x)};
    \node[color=orange] at (1.2,0.2) {$R(z)= \frac{1}{2}-\frac{1}{2} \tanh(kz)$};
\draw[color=red,dashed,samples=300] plot[id=H] function{(x<0) ? 1 : 0};
\node[color=red] at (0.8,0.9) {$R(z)=1-H(z)$};
\end{tikzpicture}
\caption{Choice of function $R$: in the limit $k\to\infty$ a
  step-function is approached, and the interaction effectively
  approaches a variant of the leader-follower model in \cite{DuMaPiWo09} (if the assertiveness of
individuals were assumed to be constant in time).}
\label{fig:choiceR}
\end{figure}
Note that in the limit $k\to\infty$ the choice \eqref{e:R} corresponds
to an approximation of $1-H(z)$ with $H$ denoting the Heaviside
function, see Figure~\ref{fig:choiceR}. In this limit, individuals are
either maximally assertive in the interaction with a value of the assertiveness
variable in $(0,1)$, or minimally assertive with assertiveness in
$(-1,0)$. Effectively, one would recover in the same limit a variant
of the opinion leader-follower model in  \cite{DuMaPiWo09} if the assertiveness of
individuals were assumed to be constant in time.

We conclude by discussing the choice of $D(\cdot).$ We assume that the ability to change individual opinions by
self-thinking decreases as one gets closer to the extremal opinions. This
reflects the fact that extremal opinions are more difficult to change.
Therefore, we assume that the {self-thinking function\/} $D(\cdot)$ is a decreasing function
of $v^2$ with $D(1)=0.$
We also need to choose the set $\mathcal{B},$ i.e.\ we have to specify
the range of values the random variables $\tilde\eta$, $\eta$ in
\eqref{e:exchange} can assume.
Clearly, it depends on
the particular choice for $D(\cdot).$ Let us consider the upper bound at
$w=1$ first. To ensure that individuals' opinions do not leave
$\mathcal{I}$, we need
\begin{align*}
 v^* = v-\gamma C(x,y,v,w)(v-w)+\tilde\eta D( v)\leq 1
\end{align*}
Obviously, the worst case is $w=1$, where we have to ensure
$$\tilde\eta D( v)\leq 1-v+\gamma (v-1)=(1-v)(1-\gamma
)$$
Hence, if $D( v)/(1-v)\leq K_+$ it suffices to have
$|\tilde\eta|\leq(1-\gamma)/K_+$.
 A similar computation for the lower boundary
shows that if $D( v)/(1+v)\leq K_-$ it suffices to have
$|\tilde\eta|\leq(1-\gamma)/K_-$. 

\medskip

In this setting, we are led to study the evolution of the distribution
function as a function depending on the assertiveness
$x\in\mathcal{J}$, opinion $w\in\mathcal{I}$ and time
$t\in\mathbb{R}^+$, $f=f(x,w,t)$. In analogy with the classical
kinetic theory of rarefied gases, we emphasise the role of the
different independent variables by identifying the velocity with
opinion, and the position with assertiveness. In this way, we assume
at once that the variation of the distribution $f(x,w,t)$ with
respect to the opinion variable $w$ depends on `collisions' between
agents, while the time change of distributions with respect to the
assertiveness $x$ depends on the transport term. Unlike in
physical applications where the transport term involves the velocity field
of particles, here the transport term contains an equivalent `opinion-velocity field' $\Phi=\Phi(x,w)$
which controls the flux depending on the independent
variables $x$ and $w$. 
This is in contrast to the physical situation of rarefied gases where the field would
simply be given by $\Phi(w)=w$.

The time-evolution of the distribution function $f = f(x,w,t)$ of individuals
depending on assertiveness $x \in \mathcal{J}$, opinion $w \in
\mathcal{I}$ and
time $t \in \mathbb{R}^+$ will obey an  {\em inhomogeneous
Boltzmann-type equation\/},
\begin{align}
\label{eq:}
\frac{\partial }{\partial t}f(x,w,t) + \Div_x(\Phi(x,w)f(x,w,t))= \frac{1}{\tau} \mathcal{Q}(f,f)(x,w,t).
\end{align}
Herein, $\Phi$ is the `opinion-velocity field' and $\tau$ is a suitable relaxation time which allows to control
the interaction frequency. 
The Boltzmann-like collision operator $\mathcal{Q}$ which describes
the change of density due to binary interactions
is derived by standard methods of kinetic theory. A useful way of writing the collision
operators is the so-called weak form.
 It corresponds to consider, for all smooth functions $\phi(w)$,
\begin{align}\label{collisional}
&\int_{\mathcal{J}}\int_{\mathcal{I}}\mathcal{Q}(f,f)(x,w,t)\phi(w)\,dw\,dx{}\\ 
=&\frac12\left\langle \int_{\mathcal{J}}\int_{\mathcal{I}^2}  \bigl(  \phi(w^*)+\phi(v^*)-\phi(w)-\phi(v)\bigr)
f(x,v,t) f(y,w,t) \,dv \, dw\,dx\right\rangle ,\nonumber
\end{align}
where $\langle \cdot \rangle$ denotes the operation of mean with respect to
the random quantities $\tilde\eta$, $\eta$.

Choosing $\phi(w)=1$ as test function in \eqref{collisional} and denoting the mass by
$$
\rho(f)(t)=\int_{\mathcal{J}}\int_{\mathcal{I}} f(x,w,t)\,dw\,dx
$$
implies conservation of mass,
$$
\frac{{\rm d}\rho(f)(t)}{{\rm d} t}=\frac{1}{\tau} \int_{\mathcal{J}}\int_{\mathcal{I}} \mathcal{Q}(f,f)(x,w,t)\,dw\,dx=0.
$$
If there is point-wise conservation of opinion in each
interaction in \eqref{e:exchange} (e.g.\ choosing $C \equiv 1$ and
$D \equiv 0$), 
$$
v^*+w^*=v+w
$$
or, more generally, conservation of opinion in the mean in each
interaction in \eqref{e:exchange} (e.g.\ choosing $C \equiv 1$),
$$
\langle v^*+w^* \rangle =v+w,
$$
then the first moment, the mean opinion, is also conserved. This can
be seen by choosing $\phi(w)=w$ in \eqref{collisional}. Denoting the mean opinion by
$$
m(f)(t)=\int_{\mathcal{J}}\int_{\mathcal{I}} f(x,w,t)w\,dw\,dx,
$$
we obtain
$$
\frac{{\rm d}m(f)(t)}{{\rm d} t}=\frac{1}{\tau} \int_{\mathcal{J}}\int_{\mathcal{I}} \mathcal{Q}(f,f)(x,w,t)w\,dw\,dx=0.
$$

We still have to specify the `opinion-velocity field' $\Phi(x,w)$. A possible choice can be
\begin{align}\label{e:bigsortphi}
\Phi(x,w)= G(x,w,t) (1-x^2)^\alpha,
\end{align}
where $G = G(x,w,t)$ models the in- or decrease of the assertiveness level, while the prefactor $(1-x^2)^{\alpha}$ ensures that
the assertiveness level stays inside the domain $\mathcal{J}$.

This leads to an {\em inhomogeneous
Boltzmann-type equation\/} of the following form
\begin{equation}
\label{inhB}
\frac{\partial}{\partial t}f+\Div_x (G(x,w,t) (1-x^2)^\alpha f)=\frac1\tau \mathcal{Q}(f,f).
\end{equation}

\subsection{A multidimensional Boltzmann-type equation}
\label{sec:multidBoltzmann}

Alternatively, we can consider that both, change of opinion and and
change of the assertiveness level happen through binary collisions.
In this case we have the following interaction rules:
\begin{subequations}
\label{e:leadership2}
\begin{align}
v^* &= v - \gamma R(x,x-y)P(|v-w|)(v-w) + \tilde\eta D(v),\\
x^* &= x + \delta \tilde G(x,x-y)P(|v-w|)(x-y) + \mu  D(x),\\
w^* &= w - \gamma R(y,y-x)P(|v-w|) (w-v) + \eta D(w),\\
y^* &= y + \delta \tilde G(y,y-x)P(|v-w|)(y-x) + \tilde\mu D(y).
\end{align}
\end{subequations}
Herein, the functions $P$, $R$, and $D$ play the same role as in the
previous section, and are assumed to fulfil the assumptions introduced
earlier.
The constant parameters $\gamma \in(0,1/2)$ and $\delta \in(0,1/2)$
control the `speed' of attraction of opinions and repulsion of
assertiveness, respectively. 
In addition, we introduce the function $\tilde G$ which describes the in- and
decrease of the individual 
assertiveness level. It is assumed to fulfil
$$
0 \leq \tilde G(x,x-y) \leq 1.
$$
Together, these assumptions guarantee that $v^*, w^*\in \mathcal{I}$  and
$y^* ,x^*\in\mathcal{J}$.
The quantities $\tilde\eta$, $\eta$ and $\tilde\mu$, $\mu$ are uncorrelated random
variables with distribution $\Theta$ with
variances $\sigma_{\eta}^2$ and $\sigma_{\mu}^2$, respectively, and
zero mean, assuming values on a set $\mathcal{B}\subset\mathbb{R}.$

To specify $\tilde G$, we make the assumption that highly assertive
individuals gain more assertiveness and weakly assertive ones loose
assertiveness in a collision. Therefore, we propose the following form:  
\begin{equation}
\label{e:Gtilde}
\tilde G(x,x-y) =  (1-x^2)^\alpha \lvert x-y\rvert,
\end{equation}
where the quadratic polynomial ensures that the assertiveness level remains in $\mathcal{J}$.
\medskip

In this setting, we are led to study the evolution of the distribution
function as a function depending on the assertiveness
$x\in\mathcal{J}$, opinion $w\in\mathcal{I}$ and time
$t\in\mathbb{R}^+$, $f=f(x,w,t)$. Different to the previous section, we assume
that the variation of the distribution $f(x,w,t)$ with
respect to both variables, assertiveness $x$ and opinion $w$, depends
on collisions between individuals.
The time-evolution of the distribution function $f = f(x,w,t)$ of individuals
depending on assertiveness $x \in \mathcal{J}$, opinion $w \in
\mathcal{I}$ and
time $t \in \mathbb{R}^+$ will obey a \textit{multi-dimensional homogeneous Boltzmann-type equation},
\begin{align}
\label{e:homB}
\frac{\partial }{\partial t}f(x,w,t) = \frac{1}{\tau} \mathcal{Q}(f,f)(x,w,t),
\end{align}
where $\tau$ is a suitable relaxation time which allows to control
the interaction frequency. 
The Boltzmann-like collision operator $\mathcal{Q}$ which describes
the change of density due to binary interactions
is derived by standard methods of kinetic theory. 
 The weak form of \eqref{e:homB} is given by
\begin{multline*}
\int_{\mathcal{J}}\int_{\mathcal{I}}\mathcal{Q}(f,f)(x,w,t)\phi(x,w)\,dw\,dx{}\\ 
=\frac12\biggl\langle \int_{\mathcal{J}^2}\int_{\mathcal{I}^2}  \bigl(
\phi(x^*,w^*)+\phi(y^*,v^*)-\phi(x,w)-\phi(y,v)\bigr) 
f(x,v,t) f(y,w,t) \,dv \, dw\,dy\,dx\biggr\rangle ,
\end{multline*}
for all test functions $\phi(x,w)$ where $\langle \cdot \rangle$ denotes the operation of mean with respect to
the random quantities $\tilde\eta$, $\eta$ and $\tilde\mu$, $\mu$.

%%%%%%%%%%%%%%%%%%%%%%%%%%%%%%%%%%%%%%%%%%%%%%%%%%%

\section{Political segregation: `The Big Sort'}
\label{sec:bigsort}

% write a bit more about the Big Sort, either here or in the introduction

In this section we are interested in modelling the clustering of individuals who
share similar political opinions, a process that has been observed in
the USA over the last decades. 

In 2008, journalist Bill Bishop
achieved popularity as the author of the book {\em The Big Sort: Why
  the Clustering of Like-Minded America Is Tearing Us Apart}
\cite{Bishop}. 
Bishop's thesis is that
US citizens increasingly choose to live among politically like-minded
neighbours. Based on county-level election results of US presidential elections in
the past thirty years, he observed a doubling of so-called
`landslide counties', counties in which either candidate won or lost
by 20 percentage points or more. Such a segregation of political
supporters may result in making political debates more bitter and hamper
the political decision-making by consensus.

Bishop's findings were discussed and acclaimed in many newspapers and
magazines, and former president Bill 
Clinton urged audiences to read the book. On the other hand his claims
also met opposition from political sociologists \cite{AbrFly12} who
argued that political segregation is only a by-product of the
correlation of political opinions with other sociologic 
(cultural background, race, ...) and economic factors
which drive citizens residential preferences. One could consider to
include such additional factors in a generalised kinetic model, and
potentially even couple the opinion dynamics with a kinetic model for
wealth distribution \cite{DuTo07,DuMaTo08}. 

`The Big Sort' lends itself as an ideal subject to study inhomogeneous kinetic
models for opinion formation. In this context we are faced with
spatial inhomogeneity rather than inhomogeneity in assertiveness as in
Section~\ref{sec:leadership}. In the following we propose a kinetic
model for opinion formation when the individuals are driven towards
others with a similar political opinion.

\subsection{An inhomogeneous Boltzmann-type equation}

We study the evolution of the distribution
function of political opinion as a function depending on three
independent variables, the continuous political opinion variable
$w\in [-1,1]$,
the position $x\in %\Omega\subset
\mathbb{R}^2$ on our (virtual) map and time
$t\in\mathbb{R}^+$, $f=f(x,w,t)$. In analogy with the classical
kinetic theory of rarefied gases, we assume
that the variation of the distribution $f(x,w,t)$ with
respect to the political opinion variable $w$ depends on collisions between
individuals, while the time change of distributions with respect to the
position $x$ depends on the transport term, which contains the `opinion-velocity field' $\Phi =\Phi(x,w)$.
The specific choice of $\Phi$ is discussed further below.

The exchange of opinion is modelled by binary collisions in the
operator $\mathcal{Q} $ and has a similar structure as \eqref{e:leadership2}.
Let $x = (x_1,x_2)$ and $y = (y_1,y_2)$ denote the positions of 
two individuals with opinion $v$ and $w$, respectively. Then the interaction
rule reads as:
\begin{subequations}\label{e:bigsort}
\begin{align}
v^* &= v - \gamma K(\lvert x-y\rvert)P(|v-w|)(v-w) + \tilde\eta D(v),\\
w^* &= w - \gamma K(\lvert y-x \rvert)P(|v-w|) (w-v) + \eta D(w).
\end{align}
\end{subequations}
In \eqref{e:bigsort} we make the following assumptions:
\begin{itemize}
\item Individuals exchange opinion if they are close to each other, i.e. their
physical distance is less than a certain radius. This is modelled by
the function $K(z) = \mathbf{1}_{\{\lvert z \rvert \leq r\}}$ with
some radius $r>0$.
\item The functions $P$ and $D$ and the random quantities
  $\tilde\eta$, $\eta$ satisfy the same assumptions as in Section~\ref{sec:leadership}.
\end{itemize}
We still need to specify the `opinion-velocity field' $\Phi(x,w)$. As as first approach it helps to think of $\Phi(x,w)$ modelling the
movement of individuals with respect to a given initial configuration:
we consider $\mathbb{R}^2$ and a number of
counties $\Omega_i$ which constitute a
disjoint cover of $\mathbb{R}^2$, each with a county seat
$c_i=(c^{(1)}_i,c^{(2)}_i)\in\mathbb{R}^2\times [-1,1]$ with position in $\mathbb{R}^2$
and an election result (by plurality voting system) which is either
`red/Republican'  ($w>0$) or `blue/Democratic' ($w<0$). In our numerical experiments presented later we will typically
restrict ourselves to bounded domains $\Omega\subset \mathbb{R}^2$ with no-flow
boundary conditions.

We assume that supporters of a party are attracted to counties which
are controlled by the party they support. We model this
effect by defining $\Phi(x,w)$ as a potential that drives the dynamics
and is given by a superposition of (signed) Gaussians $C(x)$ centered around
$c_i^{(1)}$ with variance $\sigma_i.$ We assume also that stronger supporters, i.e.\ individuals
with more extreme opinion values, are able to retain there positions.
A possible way to take these effects into account is to choose
\begin{align}\label{eq:potential}
\Phi(x,w)=\sgn(w)\nabla C(x)(1-|x|^\beta).
\end{align}

The time-evolution of the distribution function $f = f(x,w,t)$ of individuals
with political opinion $w \in\mathcal{I}$ at position $x\in \mathbb{R}^2$ at
time $t \in \mathbb{R}^+$ will obey an {\em inhomogeneous
Boltzmann-type equation\/} for the distribution function $f=f(x,w,t)$ which
is of the form
\begin{equation}
\label{inhB2}
\frac{\partial}{\partial t}f+\Div_x(\Phi(x,w) f )=\frac1\tau \mathcal{Q}(f,f).
\end{equation}
In a two-party system as we consider it, quantities of interest are
the supporters of the parties, given by the marginals
\begin{align}\label{e:RB}
f_D(x,t)=\int_{-1}^0 f(x,w,t)\;dw,\quad f_R(x,t)=\int_0^1 f(x,w,t)\;dw.
\end{align}
In simulations, elections could be run at time $t$ by computing
\begin{align}
D_i(t)&=\int_{\Omega_i} f_D (x,t)\;dx=\int_{\Omega_i}\int_{-1}^0 f(x,w,t)\;dw\,dx, \label{e:blue}\\
R_i(t)&=\int_{\Omega_i} f_R (x,t)\;dx=\int_{\Omega_i}\int_0^1 f(x,w,t)\;dw\,dx \label{e:red},
\end{align}
to adapt the values of $c^{(2)}_i$ defining the attractive Gaussians
$C(x)$ in \eqref{eq:potential}
accordingly in time.

%%%%%%%%%%%%%%%%%%%%%%%%%%%%%%%%%%%%%%%%%%%%%%%%%%%%%%%%%%%%%%%%%%%%%%

\section{Fokker-Planck limits}
\label{sec:fokkerplanck}

\subsection{Fokker-Planck limit for the inhomogeneous
  Boltzmann equation}

In general, equations like \eqref{inhB2} (and \eqref{inhB}) are rather difficult to treat and it is usual in kinetic
theory to study certain asymptotics, which
frequently lead to simplified models of Fokker-Planck type. 
To this end, we study by formal asymptotics 
the quasi-invariant opinion
limit ($\gamma,\sigma_\eta \to 0$ while keeping $\sigma_\eta^2/\gamma = \lambda$
fixed), following the path laid out in
\cite{Tosc06}.

Let us introduce some notation. First, consider test-functions
$\phi \in\mathcal{C}^{2,\delta}([-1,1])$ for some $\delta>0$. We use
the usual H\"older norms 
\begin{align*}
 \|\phi\|_{\delta} = \sum_{|\alpha|\leq 2}\|D^\alpha\phi\|_{\mathcal{C}}+\sum_{\alpha=2}[D^\alpha\phi]_{\mathcal{C}^{0,\delta}},
\end{align*}
where $[h]_{\mathcal{C}^{0,\delta}} = \sup_{v\neq w}{|h(v)-h(w)|}/{|v-w|^\delta}.$
Denoting by $\mathcal{M}_0(A)$, $A\subset\mathbb{R}$ the space of probability measures on $A$, we define
\begin{align*}
 \mathcal{M}_p(A)=\left\{\Theta\in\mathcal{M}_0\;\left|\;\int_A |\eta|^pd\Theta(\eta)<\infty,\,p\geq 0\right.\right\},
\end{align*}
the space of measures with finite $p$th momentum. In the following all our probability densities belong to $\mathcal{M}_{2+\delta}$ and we assume that the density $\Theta$ is obtained from a random variable $Y$ with zero mean and unit variance. We then obtain
\begin{align}\label{e:expvalue}
 \int_{\mathcal{I}}|\eta|^p\Theta(\eta)\;d\eta = {\mathrm E}[|\sigma_\eta  Y|^p]=\sigma_\eta^p{\mathrm E}[|Y|^p],
\end{align}
where ${\mathrm E}[|Y|^p]$ is finite.
The weak form of \eqref{inhB2} is given by
\begin{multline}
\label{weakform}
\frac{d}{dt}\int_{\mathcal{I}\times \mathbb{R}^2} f(x,w,t) \phi(w)\,dw\,dx +\int_{\mathcal{I}\times \mathbb{R}^2}
\Div_x(\Phi(x,w) f(x,w,t) )\phi(w)\,dw\,dx\\
=\frac1{\tau}\int_{\mathcal{I}\times \mathbb{R}^2}
\mathcal{Q}(f,f)(w)\phi(w)\,dw\,dx
\end{multline}
where 
\begin{align*}
\begin{split}
\int_{\mathcal{I}\times \mathbb{R}^2}&\mathcal{Q}(f,f)(w)\phi(w)\,dw\,dx {}\\
&=\frac12\left\langle \int_{\mathcal{I}^2 \times \mathbb{R}^2}  \bigl(  \phi(w^*)+\phi(v^*)-\phi(w)-\phi(v)\bigr)
f(x,v) f(x,w) \,dv \, dw\,dx\right\rangle .
\end{split}
\end{align*}
To study the situation for large times, i.e.\ close to the steady
state, we introduce for $\gamma\ll 1$ the transformation
\begin{equation*}
\tilde t=\gamma t, \;\tilde x=\gamma x,\; g(\tilde x,w,\tilde t )=f(x,w,t).
\end{equation*}
This implies $f(x,w,0)=g(\tilde x,w,0)$ and the evolution of the scaled
density $g(\tilde x,w,\tilde t)$ follows (we immediately drop the tilde in the
following and denote the rescaled variables simply by $t$ and $x$)
\begin{multline}
\label{weakform2}
\frac{d}{dt}\int_{\mathcal{I}\times \mathbb{R}^2} g(x,w,t)
\phi(w)\,dw\,dx +\int_{\mathcal{I}\times \mathbb{R}^2}
\Div_x(\Phi(x,w) g(x,w,t)) \phi(w)\,dw\,dx\\=\frac1{\gamma\tau}\int_{\mathcal{I}\times \mathbb{R}^2}
\mathcal{Q}(g,g)(w)\phi(w)\,dw\,dx.
\end{multline}
Due to the collision rule \eqref{e:exchange}, it holds
\begin{align*}
  w^*-w &= -\gamma C(x,y,v,w) (w-v)+\eta D(w)\ll 1.
\end{align*}
Taylor expansion of $\phi$ up to second order around $w$ of the right
hand side of \eqref{weakform2} leads to
\begin{align*}
&\Big\langle \frac{1}{\gamma\tau}\int_{\mathcal{I}^2 \times \mathbb{R}^2}\phi'(w)\left[-\gamma C(x,y,v,w)(w-v)+\eta D(w)\right]g(x,w)g(x,v)\,dv\,dw\,dx\Big\rangle\\
&+\Big\langle \frac{1}{2\gamma\tau}\int_{\mathcal{I}^2 \times \mathbb{R}^2}\phi''(\tilde{w})\left[-\gamma C(x,y,v,w)(w-v)+\eta D(w)\right]^2g(x,w)g(x,v)\,dv\,dw\,dx\Big\rangle\\
=&\frac{1}{\gamma{\tau}}\int_{\mathcal{I}^2 \times \mathbb{R}^2}\phi'(w)\left[-\gamma C(x,y,v,w)(w-v))\right]g(x,w)g(x,v)\,dv\,dw\,dx\\
&+\Big \langle \frac{1}{2\gamma\tau}\int_{\mathcal{I}^2 \times \mathbb{R}^2}\phi''({w})\big[\gamma
C(x,y,v,w)(w-v)+\eta D(w)\big]^2g(x,w)g(x,v)\,dv\,dw\,dx\Big \rangle +R(\gamma,\sigma_\eta)\\
=&-\frac 1{\tau}\int_{\mathcal{I}\times \mathbb{R}^2}\phi'(w)\mathcal{K}(x,w)g(x,w)\,dw\,dx\\
&+ \frac{1}{2\gamma\tau}\int_{\mathcal{I}^2 \times \mathbb{R}^2}\phi''(w)\Big[\gamma^2
C^2(x,y,v,w)(w-v)^2+\gamma\lambda D^2(w)\Big]g(x,w)g(x,v)\,dv\,dw\,dx+R(\gamma,\sigma_\eta),
\end{align*}
with $\tilde{w}=\kappa w^*+(1-\kappa)w$ for some $\kappa\in [0,1]$ and
\begin{multline*}
R(\gamma,\sigma_\eta)=\Big\langle
\frac{1}{2\gamma\tau}\int_{\mathcal{I}^2 \times \mathbb{R}^2}(\phi''(\tilde{w})-\phi''(w)) \\
\times \left[-\gamma C(x,y,v,w)(w-v)+\eta D(w)\right]^2g(x,w)g(x,v)\,dv\,dw\,dx\Big\rangle
\end{multline*}
and
\begin{equation*}
\mathcal{K}(x,w) = \int_{\mathcal{I}} C(x,y,v,w)(w-v) g(x,v)\,dv.
\end{equation*}
Now we consider the limit $\gamma, \sigma_\eta  \to 0$ while keeping $\lambda=\sigma_\eta^2/\gamma$
fixed. 

We first show that the remainder term $R(\gamma,\sigma_\eta)$ vanishes is this limit, similar as in
\cite{Tosc06}. 
 Note first that as $\phi\in\mathcal{F}_{2+\delta}$, by the collision
 rule \eqref{e:exchange} and the definition of $\tilde{w}$ we have
\begin{align*}
|\phi''(\tilde{w})-\phi''(w)| \leq \|\phi''\|_\delta |\tilde{w}-w|^\delta \leq\|\phi''\|_\delta |w^*-w|^{\delta} 
=\|\phi''\|_\delta \left|\gamma C(x,y,v,w)(w-v)+\eta D(w)\right|^\delta.
\end{align*}
Thus we obtain
\begin{equation*}
R(\gamma,\sigma_\eta )\leq
\frac{\|\phi''\|_\delta}{2\gamma\tau}\Big\langle\int_{\mathcal{I}^2 \times \mathbb{R}^2}\left[-\gamma
  C(x,y,v,w)(w-v)+\eta D(w)\right]^{2+\delta} g(x,w)g(x,v)\,dv\,dw\,dx\Big\rangle.
\end{equation*}
Furthermore, we note that
\begin{align*}
&\left[\eta D(w)-\gamma C(x,y,v,w)(w-v)\right]^{2+\delta} 
\leq 2^{1+\delta}\left(|\gamma C(x,y,v,w)(w-v)|^{2+\delta}+|\eta D(w)|^{2+\delta}\right)\\
\leq&  2^{3+2\delta}|\gamma|^{2+\delta}+2^{1+\delta}|\eta |^{2+\delta}.
\end{align*}
Here, we used the convexity of $f(s):=|s|^{2+\delta}$ and the fact that $w,v\in\mathcal{I}$ and thus bounded. We conclude
\begin{align*}
|R(\gamma,\sigma_\eta )|\leq \frac{C\|\phi''\|_\delta}{\tau}\left(\gamma^{1+\delta}+\frac{1}{2\gamma}\left<|\eta |^{2+\delta}\right>\right)
= \frac{C\|\phi''\|_\delta}{\tau}\left(\gamma^{1+\delta}+\frac{1}{2\gamma}\int_{\mathcal{I}}|\eta|^{2+\delta}\Theta(\eta)\;d\eta\right).
\end{align*}
Since $\Theta\in\mathcal{M}_{2+\delta}$, and $\eta$ has variance $\sigma_\eta^2$ we have (see \eqref{e:expvalue})
\begin{align}
\int_{\mathcal{I}}|\eta
|^{2+\delta}\Theta(\eta)\,d\eta={\mathrm E}\left[\left|\sqrt{\lambda\gamma}Y\right|^{2+\delta}\right]=(\lambda\gamma)^{1+\frac{\delta}{2}}{\mathrm E}\left[\left|Y\right|^{2+\delta}\right]. 
\end{align}
Thus, we conclude that the term $R(\gamma,\sigma_\eta )$ vanishes in the limit $\gamma,
\sigma_\eta \to 0$ while keeping $\lambda=\sigma_\eta^2/\gamma$ fixed.

Then, in the same limit, the term on the right
hand side of \eqref{weakform2} converges to
\begin{multline*}
-\frac{1}{\tau}\int_{\mathcal{I}\times
  \mathbb{R}^2}\phi'(w)\mathcal{K}(x,w)g(x,w)\,dw\,dx
+\frac{1}{2\tau}\int_{\mathcal{I}^2
  \times \mathbb{R}^2}\phi''(w)\left[\lambda
  D^2(w)\right]g(x,w)g(x,v)\,dv\,dw\,dx\\
=-\frac 1{\tau}\int_{\mathcal{I}\times
  \mathbb{R}^2}\phi'(w)\mathcal{K}(x,w)g(x,w)\,dw\,dx
+\frac{\lambda}{2\tau}\int_{\mathcal{I}\times \mathbb{R}^2}\phi''(w)
M(x)D^2(w)g(x,w)\,dw\,dx,
\end{multline*}
with $M(x) = \int g(x,v) \,dv$ being the mass of individuals in $x$.
After integration by parts we obtain the right hand side of (the weak form of) the Fokker-Planck equation
\begin{multline}
\label{e:fokkerplanck}
\frac{\partial}{\partial \tau}g(x,w,t)+ \Div_x(\Phi(x,w) g(x,w,t))\\=\frac{\partial}{\partial w}\left(\frac 1{\tau}\mathcal{K}(x,w,t)g(x,w,t)\right){}
+\frac{\lambda M(x)}{2\tau}\frac{\partial^2}{\partial w^2}\left(D^2(w)g(x,w,t) \right),
\end{multline}
subject to no flux boundary conditions for the variable $w$ (which result from the integration by parts).

\subsection{Fokker-Planck limit for the multidimensional model}

We follow a similar approach as in the previous section, now in the
two-dimensional setting of the opinion formation model
\eqref{e:leadership2} (cf.\ also \cite{ParTos14}).
Our aim is to study by formal asymptotics 
the quasi-invariant opinion
limit where $\gamma,\delta,\sigma_\eta,\sigma_\mu \to 0$ while keeping $\sigma_\eta^2/\gamma =c_1\sigma_\mu^2/(c_2\delta)=\lambda$ fixed, introducing the positive constants
$c_1=\delta/\gamma$ and $c_2=\sigma_\mu/\sigma_\eta$.

We consider test-functions
$\phi \in\mathcal{C}^{2,\delta}([-1,1]^2)$ for some $\delta>0$. As
above we use the usual H\"older norms and consider
denote by $\mathcal{M}_p(A)$, $A\subset\mathbb{R}$ the space of
probability measures on $A$ with finite $p$th momentum. In the following all our probability densities belong to $\mathcal{M}_{2+\delta}$ and we assume that the density $\Theta$ is obtained from a random variable $Y$ with zero mean and unit variance. We then obtain
\begin{align*}
 \int_{\mathcal{I}}|\eta|^p\Theta(\eta)\;d\eta = {\mathrm E}[|\sigma
 Y|^p]=\sigma_\eta^p{\mathrm E}[|Y|^p],\quad
 \int_{\mathcal{I}}|\mu|^p\Theta(\mu)\;d\mu = {\mathrm E}[|\tilde\sigma Y|^p]=\sigma_\mu^p{\mathrm E}[|Y|^p],
\end{align*}
where ${\mathrm E}[|Y|^p]$ is finite.
The weak form of \eqref{e:homB} is given by
\begin{equation}
\label{weakformhomB}
\frac{d}{dt}\int_{\mathcal{J}\times\mathcal{I}} f(x,w,t) \phi(x,w)\,dw\,dx 
=\frac1{\tau}\int_{\mathcal{J}\times\mathcal{I}}
\mathcal{Q}(f,f)(x,w)\phi(x,w)\,dw\,dx
\end{equation}
where 
\begin{multline*}
\int_{\mathcal{J}\times\mathcal{I}}\mathcal{Q}(f,f)(x,w)\phi(x,w)\,dw\,dx {}\\
=\frac12\biggl\langle \int_{\mathcal{J}^2 \times \mathcal{I}^2}  \bigl(  \phi(x^*,w^*)+\phi(y^*,v^*)-\phi(x,w)-\phi(y,v)\bigr) f(x,v) f(x,w) \,dv \,dy \, dw\,dx\biggr\rangle .
\end{multline*}
To study the situation for large times, i.e.\ close to the steady
state, we introduce for $\gamma\ll 1$ the transformation
\begin{equation*}
\tilde t=\gamma t,\; g(x,w,\tilde t )=f(x,w,t).
\end{equation*}
This implies $f(x,w,0)=g(x,w,0)$ and the evolution of the scaled
density $g(x,w,\tilde t)$ follows  (we immediately drop the tilde in the
following and denote the rescaled time simply by $t$)
\begin{equation}
\label{weakform2homB}
\frac{d}{dt}\int_{\mathcal{J}\times\mathcal{I}}g(x,w,t)
\phi(x,w)\,dw\,dx=\frac1{\gamma\tau}\int_{\mathcal{J}\times\mathcal{I}}
\mathcal{Q}(g,g)(x,w)\phi(x,w)\,dw\,dx.
\end{equation}
Due to the collision rules \eqref{e:leadership2}, it holds
\begin{align*}
  w^*-w &= -\gamma R(x-y)P(|w-v|)(w-v)+\eta D(w)\ll 1,\\
x^*-x &= \delta \tilde G(x-y)P(|v-w|)(x-y) + \mu D(x) \ll 1 .
\end{align*}
We now employ multidimensional Taylor expansion of $\phi$ up to second order around
$(x,w)$ of the right 
hand side of \eqref{weakform2homB}. Recalling that the random
quantities $\eta$, $\mu$ have mean zero, variance $\sigma_\eta$ and
$\sigma_\mu$, respectively, and are uncorrelated, we can follow along the lines of the
computations of the previous subsection, to obtain
\begin{multline*}
\Big\langle \frac{1}{\gamma\tau}\int_{\mathcal{J}^2 \times
  \mathcal{I}^2}\frac{\partial\phi}{\partial w}(x,w)\left[-\gamma
  R(x-y)P(|w-v|)(w-v)+\eta
  D(w)\right]{ g(x,w)g(x,v)\,dv\,dy\,dw\,dx\Big\rangle}\\
+\Big\langle \frac{1}{\gamma\tau}\int_{\mathcal{J}^2 \times
  \mathcal{I}^2}\frac{\partial\phi}{\partial x}(x,w)\left[\delta \tilde G(x-y)P(|v-w|)(x-y) + \mu D(x) \right]\\
\shoveright{\times g(x,w)g(x,v)\,dv\,dy\,dw\,dx\Big\rangle}\\
+\Big\langle \frac{1}{2\gamma\tau}\int_{\mathcal{J}^2 \times
  \mathcal{I}^2}\frac{\partial^2\phi}{\partial w^2}(x,w)\left[-\gamma
  R(x-y)P(|w-v|)(w-v)+\eta
  D(w)\right]^2 \\
\shoveright{\times g(x,w)g(x,v)\,dv\,dy\,dw\,dx\Big\rangle}\\
+\Big\langle \frac{1}{2\gamma\tau}\int_{\mathcal{J}^2 \times
  \mathcal{I}^2}\frac{\partial^2\phi}{\partial x^2}(x,w)\left[\delta \tilde G(x-y)P(|v-w|)(x-y) + \mu D(x)\right]^2 \\
\shoveright{\times g(x,w)g(x,v)\,dv\,dy\,dw\,dx\Big\rangle}\\
+\Big\langle \frac{1}{2\gamma\tau}\int_{\mathcal{J}^2 \times
  \mathcal{I}^2}\frac{\partial^2\phi}{\partial w\partial x}(x,w)\left[-\gamma
  R(x-y)P(|w-v|)(w-v)+\eta D(w)\right]\\
\shoveright{\times\left[\delta \tilde G(x-y)P(|v-w|)(x-y) + \mu D(x) \right]g(x,w)g(x,v)\,dv\,dy\,dw\,dx\Big\rangle}\\
 \shoveright{+R(\gamma,\sigma_\eta ,\sigma_\mu)}\\
{=-\frac 1{\tau}\int_{\mathcal{I}\times
  \mathcal{J}^2}\frac{\partial\phi}{\partial w}(x,w)\mathcal{K}(x,w)g(x,w)\,dy\,dw\,dx}
-\frac {\delta}{\gamma\tau}\int_{\mathcal{I}^2\times\mathcal{J}}\frac{\partial\phi}{\partial x}(x,w)\mathcal{L}(x,w)g(x,w)\,dv\,dw\,dx\\
+ \frac{1}{2\gamma\tau}\int_{\mathcal{I}^2 \times \mathcal{J}^2}\frac{\partial^2\phi}{\partial w^2}(x,w)\Big[\gamma^2
R^2(x-y)P^2(|w-v|)(w-v)^2+\sigma_\eta^2D^2(w)\Big]\\
\shoveright{\times g(x,w)g(x,v)\,dv\,dy\,dw\,dx}\\
+ \frac{1}{2\gamma\tau}\int_{\mathcal{I}^2 \times \mathcal{J}^2}\frac{\partial^2\phi}{\partial x^2}(x,w)\Big[\delta^2 \tilde G^2(x-y)P^2(|v-w|)(x-y)^2 + \sigma_\mu^2 D^2(x)\Big]\\
\shoveright{\times g(x,w)g(x,v)\,dv\,dy\,dw\,dx}\\
+ \frac{1}{2\gamma\tau}\int_{\mathcal{I}^2 \times
  \mathcal{J}^2}\frac{\partial^2\phi}{\partial w\partial x}(x,w)\Big[\gamma\delta R(x-y)P^2(\lvert w-v\rvert)(w-v)\tilde G(x-y)(x-y)\Big]\\
\shoveright{\times g(x,w)g(x,v)\,dv\,dy\,dw\,dx}\\
+R(\gamma,\sigma_\eta ,\sigma_\mu),
\end{multline*}
where
\begin{subequations}\label{e:nonlocalops}
\begin{align}
\mathcal{K}(x,w) &= \int_{\mathcal{I}} R(x-y)P(\lvert w-v\rvert)(w-v)
g(x,v)\,dv,\\
\mathcal{L}(x,w) &= \int_{\mathcal{J}}\tilde G(x-y)P(|v-w|)(x-y)g(x,v)\,dy.
\end{align}
\end{subequations}
and $R(\gamma,\sigma_\eta ,\sigma_\mu)$ is the remainder term. Our aim is
now to consider
the formal limit  $\gamma,\delta,\sigma_\eta,\sigma_\mu \to 0$ while keeping $\sigma_\eta^2/\gamma =c_1\sigma_\mu^2/(c_2\delta)
=\lambda$ fixed, recalling $c_1=\delta/\gamma$ and $c_2=\sigma_\mu/\sigma_\eta$.
The remainder term $R(\gamma,\sigma_\eta ,\sigma_\mu)$ depends on the
higher moments of the (uncorrelated) random quantities and can be shown
to vanish in this limit, similar as in the previous subsection (we omit the details).
In the same limit, the term on the right
hand side of \eqref{weakform2homB} then converges to
\begin{multline*}
-\frac{1}{\tau}\int_{\mathcal{I}\times
  \mathcal{J}^2}\frac{\partial\phi}{\partial
  w}(x,w)\mathcal{K}(x,w)g(x,w)\,dy\,dw\,dx
 -\frac {c_1}{\tau}\int_{\mathcal{I}^2\times\mathcal{J}}\frac{\partial\phi}{\partial x}(x,w)\mathcal{L}(x,w)g(x,w)\,dv\,dw\,dx\\
+\frac{1}{2\tau}\int_{\mathcal{I}^2
  \times \mathcal{J}^2}\frac{\partial^2\phi}{\partial w^2}(x,w)\lambda
  D^2(w)g(x,w)g(x,v)\,dv\,dy\,dw\,dx\\
\shoveright{+\frac{1}{2\tau}\int_{\mathcal{I}^2
  \times \mathcal{J}^2}\frac{\partial^2\phi}{\partial x^2}(x,w) c_2\lambda
  D^2(x)g(x,w)g(x,v)\,dv\,dy\,dw\,dx}\\
{=-\frac{1}{\tau}\int_{\mathcal{I}\times
  \mathcal{J}^2}\frac{\partial\phi}{\partial
  w}(x,w)\mathcal{K}(x,w)g(x,w)\,dy\,dw\,dx}
 -\frac {c_1}{\tau}\int_{\mathcal{I}^2\times\mathcal{J}}\frac{\partial\phi}{\partial x}(x,w)\mathcal{L}(x,w)g(x,w)\,dv\,dw\,dx\\
+\frac{\lambda}{2\tau}\int_{\mathcal{I}\times \mathcal{J}^2}\frac{\partial^2\phi}{\partial w^2}(x,w)
M(x)D^2(w)g(x,w)\,dy\,dw\,dx\\
+\frac{c_2\lambda}{2\tau}\int_{\mathcal{I}\times \mathcal{J}^2}\frac{\partial^2\phi}{\partial x^2}(x,w)
M(x)D^2(x)g(x,w)\,dy\,dw\,dx,
\end{multline*}
with $M(x) = \int g(x,v) \,dv$ being the mass of individuals in $x$.
After integration by parts we obtain the right hand side of (the weak form of) the Fokker-Planck equation
%\begin{multiline}
\begin{align}\label{e:fokkerplanck2}
\begin{split}
\frac{\partial}{\partial \tau}g(x,w,t) &=\frac 1{\tau}\frac{\partial}{\partial w}\left(\mathcal{K}(x,w,t)g(x,w,t)\right)+
\frac {c_1}{\tau}\frac{\partial}{\partial x}\left(\mathcal{L}(x,w,t)g(x,w,t)\right){}\\
&+\frac{\lambda M(x)}{2\tau}\frac{\partial^2}{\partial w^2}\left(D^2(w)g(x,w,t) \right) +\frac{c_2\lambda M(x)}{2\tau}\frac{\partial^2}{\partial x^2}\left(D^2(x)g(x,w,t) \right),
\end{split}
\end{align}
%\end{multline}
subject to no flux boundary conditions which result from the
integration by parts, with the nonlocal operators $\mathcal{K}$, $\mathcal{L}$ given in \eqref{e:nonlocalops}.

%%%%%%%%%%%%%%%%%%%%%%%%%%%%%%%%%%%%%%%%%%%%%%%%%%%%%%%%%%%%%%%%%%%%%%%%%%

\section{Numerical experiments}
\label{sec:numerics}

In this section, we illustrate the behaviour of the different kinetic
models and the limiting Fokker-Planck-type equations with various
simulations.
We first discuss the numerical discretisation of the different Boltzmann-type equations and the corresponding Fokker-Planck-type
equations, and then present results of numerical experiments. 

While the multi-dimensional Boltzmann-type equation \eqref{e:homB} can be
solved by a classical kinetic Monte Carlo method, the Monte Carlo
simulations for the inhomogeneous Boltzmann-type equations \eqref{eq:inhB0} are more involved. On the macroscopic level the high-dimensionality
poses a significant challenge -- hence we propose a time splitting as well as a finite element discretisation with mass lumping in space.

\subsection{Monte Carlo simulations for the multi-dimensional
  Boltzmann equation}

\label{sec:MC}

We perform a series of kinetic Monte Carlo simulations for the
Boltzmann-type models presented in
Sections~\ref{sec:leadership}.\ref{sec:inhomogBoltzmann} and
\ref{sec:leadership}.\ref{sec:multidBoltzmann}.
In this kind of simulation, known as direct simulation
Monte Carlo (DSMC) or Bird's scheme, pairs of individuals are randomly
and non-exclusively selected for binary collisions, and exchange
opinion (and assertiveness in the multi-dimensional model from Section~\ref{sec:leadership}.\ref{sec:multidBoltzmann}),
according to the relevant interaction rule. 

In each simulation we consider $N=5000$ individuals, which are uniformly distributed in $\mathcal{I}\times \mathcal{J}$ at
time $t=0$. One time step in our simulation corresponds to $N$
interactions. The average steady state opinion distribution $f = f(x,w,t)$ is calculated using $M=10$ realisations. To compute 
a good approximation of the steady state, each realisation is carried out for $n=2 \times 10^6$ time steps, then the particle distribution is averaged over
$5000$ time steps. The random variables are chosen such that $\eta_i$ and $\mu_i$
assumes only values $\nu = \pm 0.02$ with equal probability. We assume
that the diffusion has the form $D(w) = (1-w^2)^\alpha$
to ensure that the opinion $w$ remains inside the interval
$\mathcal{I}$. The parameter $\alpha$ is set to $\alpha=2$ if not
mentioned otherwise.

\subsection{Probabilistic Monte Carlo simulations for the inhomogeneous Boltzmann equation}
\label{sec:pMC}

The Monte Carlo simulations for the inhomogeneous Boltzmann-type equations \eqref{eq:inhB0} are more involved.
In equation \eqref{eq:inhB0} the advection term is of conservative form, hence the transportation step cannot be translated directly to the
particle simulation. Pareschi and Seaid \cite{ParSea04} as well as
Herty {\em et al.}\cite{HePaSe06} propose a Monte Carlo method, that is based
on a relaxation approximation of conservation laws. It corresponds to a semilinear system with linear characteristic variables.
We shall briefly review the underlying idea for the conservative
transportation operator in \eqref{eq:inhB0} in the following. 

Let us consider
the linear conservation law in one spatial dimension for the function $\rho = \rho(x,t)$ with a given flux function $b(x,t): \mathbb{R} \times [0,T] \rightarrow \mathbb{R}$:
\begin{align}\label{e:con}
\partial_t \rho + \partial_x (b(x,t) \rho) = 0, ~\rho(x,0) = \rho_0(x).
\end{align}
Note that \eqref{e:con} corresponds to the transportation step in a
splitting scheme in a Monte Carlo simulation for \eqref{eq:inhB0}. Equation \eqref{e:con} can be approximated by the following semilinear relaxation system:
\begin{align}\label{e:relax}
\partial_t \rho + \partial_x v  = 0,\quad
\partial_t v  + \bar{b} \partial_x u =  - \frac{1}{\varepsilon}(v - b(x,t) \rho)
\end{align}
 with initial conditions $\rho(x,0) = \rho_0(x)$ and $v(x,0) = b(x,0)
 \rho_0(x)$ and $\bar{b}, \varepsilon>0$. The function $v$ approaches the solution at
 local equilibrium $v = \varphi(u)$, if $-\bar{b} \leq b(x,t) \leq \bar{b} \text{ for all } x $.
Note that system \eqref{e:relax} has two characteristic variables given by $v \pm \sqrt{\bar{b}}$, which correspond to particles either moving to the
left or the right with speed $\sqrt{\bar{b}}$. Hence, we introduce the kinetic variables $p$ and $q$ with
\begin{align*}
\rho = p+q \text{ and } v = \bar{b}(p-q).
\end{align*}
Then the relaxation system can be written as follows:
\begin{align}\label{e:pq}
\partial_t p  + \partial_x\big(\sqrt{\bar{b}} p\big) = \frac{1}{\varepsilon} \Big(\frac{q-p}{2} + \frac{\varphi(p+q)}{2\sqrt{\bar{b}}}\Big),\quad
\partial_t q - \partial_x\big(\sqrt{\bar{b}} q\big)  = \frac{1}{\varepsilon}\Big (\frac{p-q}{2}- \frac{\varphi(p+q)}{2 \sqrt{\bar{b}}}\Big).
\end{align}
The numerical solver is based on a splitting algorithm for system \eqref{e:pq}. It corresponds to first solving the transportation problem and then
the relaxation step. While the transportation step is straight forward, the relaxation step can be interpreted as the
evolution of a probability density. Since
\begin{align*}
p \geq 0,~~ q \geq 0,~~\frac{p}{\rho} + \frac{q}{\rho} = 1 \text { and } \partial_t \rho = 0
\end{align*}
 in the relaxation step we can calculate the solution explicitly and obtain
\begin{align*}
p = \frac{1}{2} e^{\frac{t}{\varepsilon}} + \frac{1}{2\bar{b}} b(x,t)\rho\quad   \text{ and } \quad q = -\frac{1}{2} e^{\frac{t}{\varepsilon}} - \frac{1}{2\bar{b}} b(x,t).
\end{align*}
Then the solution at the time $t^{n+1} = (n+1) \Delta t$ reads as:
\begin{align*}
p(x,t^{n+1}) = (1-\lambda) p(x,t^n) + \lambda b(x,t^n) \rho(x,t^n) ,\quad
q(x,t^{n+1}) = (1-\lambda) q(x,t^n) - \lambda b(x,t^n) \rho(x,t^n),
\end{align*}
with $\lambda = 1-e^{-{\Delta t}/{\varepsilon}}$. Let $p^n = p(x,t^n)$, $q^n = q(x,t^n)$ and $\rho^n = p^n + q^n$. Since $0 \leq \lambda \leq 1$ we can define the probability density
\begin{align*}
P^n(\xi) = \begin{cases}
{p^n}/{\rho^n} &\text{ if } \xi = \sqrt{\bar{b}},\\
{q^n}/{\rho^n} &\text{ if } \xi = -\sqrt{\bar{b}},\\
0 & \text{ elsewhere.}
\end{cases}
\end{align*}
Since $0 \leq P^n(\xi) \leq 1$ and $\sum_{\xi} P^n(\xi) = 1$, the relaxation step can be interpreted as the evolution of a probability function
\begin{align*}
P^{n+1}(\xi) =(1-\lambda) P^n(\xi) + \lambda E^n(\xi),
\end{align*} 
where $E^n(\xi) = b(x,t^n) \rho^n$ if $\xi = \sqrt{\bar{b}}$ and $E^n(\xi) = -b(x,t^n) \rho^n$ if $\xi = -\sqrt{\bar{b}}$. 
So in the probabilistic Monte Carlo method a particle either moves to the right or the left with maximum speed $\sqrt{\bar{b}}$ in the transportation step.
Then the two groups are resampled according to their distribution with respect to the equilibrium solution $b(x,t^n)\rho^n$ in the relaxation step. 
For further details we refer to \cite{ParSea04}.

\subsection{Fokker-Planck simulations}\label{s:simfp}
The discretisation of the Fokker-Planck-type equation \eqref{e:fokkerplanck} is based on a time splitting algorithm. The splitting strategy allows us to consider the interactions in the opinion variable,
i.e. the right hand side of equation \eqref{e:fokkerplanck}, and the transport step in space separately.\\
Let $\Delta t$ denote the size of each time step and  $t^k = k \Delta t$. Then the splitting scheme consists of a transport step 
$S^1(g,\Delta t)$  for a small time interval $\Delta t$:
\begin{subequations}\label{e:transportstep}
\begin{align}
\frac{\partial g^*}{\partial t}(x,w,t) + \Div_x \big (\phi(x,w) g^*(x,w,t) \big)&= 0,\\
g^*(x,w,0) &= g^*_0(x,w),
\end{align}
\end{subequations}
and an interaction step $S^2(g,\Delta t)$:
\begin{subequations}\label{e:interactionstep}
\begin{align}
 \frac{\partial g^\diamond}{\partial t}(x,w,t) &= \frac{\partial }{\partial w}\Big(\frac{1}{\tau}\mathcal{K}(x,w,t) g^\diamond(x,w,t) \Big) + \frac{\lambda M(x)}{2\tau} \frac{\partial^2}{\partial w^2}\big(D^2(w) g^\diamond(x,w,t)\big),\\
g^\diamond(x,w,0) &= g^*(x,w,\Delta t).
\end{align}
\end{subequations}
The approximate solution at time $t=t^{k+1}$ is given by
\begin{align*}
%g^k \rightarrow g^{\diamond,k+\frac{1}{2}} \rightarrow g^{*, k+1} \rightarrow g^{\diamond, k+1}
g^{k+1}(x,w) = S_2(g^{*,k+1}, \Delta t/ 2) \circ S_1 (g^{\diamond,k+\frac{1}{2}}, \Delta t) \circ S_2(g^k,\Delta t / 2),
\end{align*}
where the superscript indizes denote the solution of $g^{\diamond}$ and $g^*$ at the discrete time steps $t^k = k \Delta t$ and $t^{k+\frac{1}{2}} = (k+\frac{1}{2}) \Delta t$.
Both equations are solved using an explicit Euler scheme in time and a conforming finite element discretisation with linear basis functions $p_i, i=1, \ldots (\#\text{points})$, in space as well as opinion. 
Then the discrete transportation step \eqref{e:transportstep} in space has the form
\begin{align*}
\mathbf{M}\Big[g^{k+1}_l(x,\cdot)-g^k_l(x,\cdot) \Big] = \Delta t \Big(\mathbf{T}\big[g^k_l(x,\cdot)\big]\Big),
\end{align*}
where $\mathbf{M} = \langle p_i, p_j \rangle_{ij}$ corresponds to the
mass matrix for element-wise linear basis functions and $\mathbf{T} =
\langle \Phi(x) \nabla p_i, p_j \rangle_{ij}$ denotes the matrix corresponding to the convective field $\Phi$ of the form \eqref{e:bigsortphi}.\\ 
In the interaction step we approximate the mass matrix by the corresponding lumped mass matrix $\tilde{\mathbf{M}}$, which can be
inverted explicitly. The discrete interaction step \eqref{e:interactionstep} reads as: 
\begin{align*}
\tilde{\mathbf{M}}\Big [g^{k+1}_l(\cdot,w)-g^k_l(\cdot,w) \Big] = \Delta t \Big(\frac{1}{\tau} \mathbf{C}\big[K[g^k_l], g^k_l(\cdot,w)\big] + \frac{\lambda M(\cdot)}{2\tau} \mathbf{L}\big[g^k_l(\cdot,w)\big]\Big),
\end{align*}
with the discrete convolution-transportation matrix $\mathbf{C} = \langle K(x,w) p_i, \nabla p_j \rangle $ and Laplacian $\mathbf{L} = \langle D(w) \nabla p_i, \nabla p_j \rangle$. Here the vector $K$ corresponds to the
discrete convolution operator $\mathcal{K}$, which has been
approximated by the midpoint rule:
\begin{align*}
K(x,w_l) = \sum_k C(x,v_k, w_l)(w_l - v_k) g(x,v_k) \Delta w. 
\end{align*}

\subsection{Numerical results: opinion formation and opinion leadership}

We present numerical results for the kinetic models for opinion
formation including the assertiveness of individuals from
Section~\ref{sec:leadership}. We consider the inhomogeneous
Boltzmann-type model of
Section~\ref{sec:leadership}.\ref{sec:inhomogBoltzmann} and the
multi-dimensional Boltzmann-type equation of
Section~\ref{sec:leadership}.\ref{sec:multidBoltzmann} which are
discretised using the Monte Carlo methods presented in the previous sections.

\subsubsection{Example 1: Influence of the interaction radius $r$}
\begin{figure}[h!]
\begin{center}
\subfigure[Inhomog. $r=1$]{\includegraphics[width=0.475\textwidth]{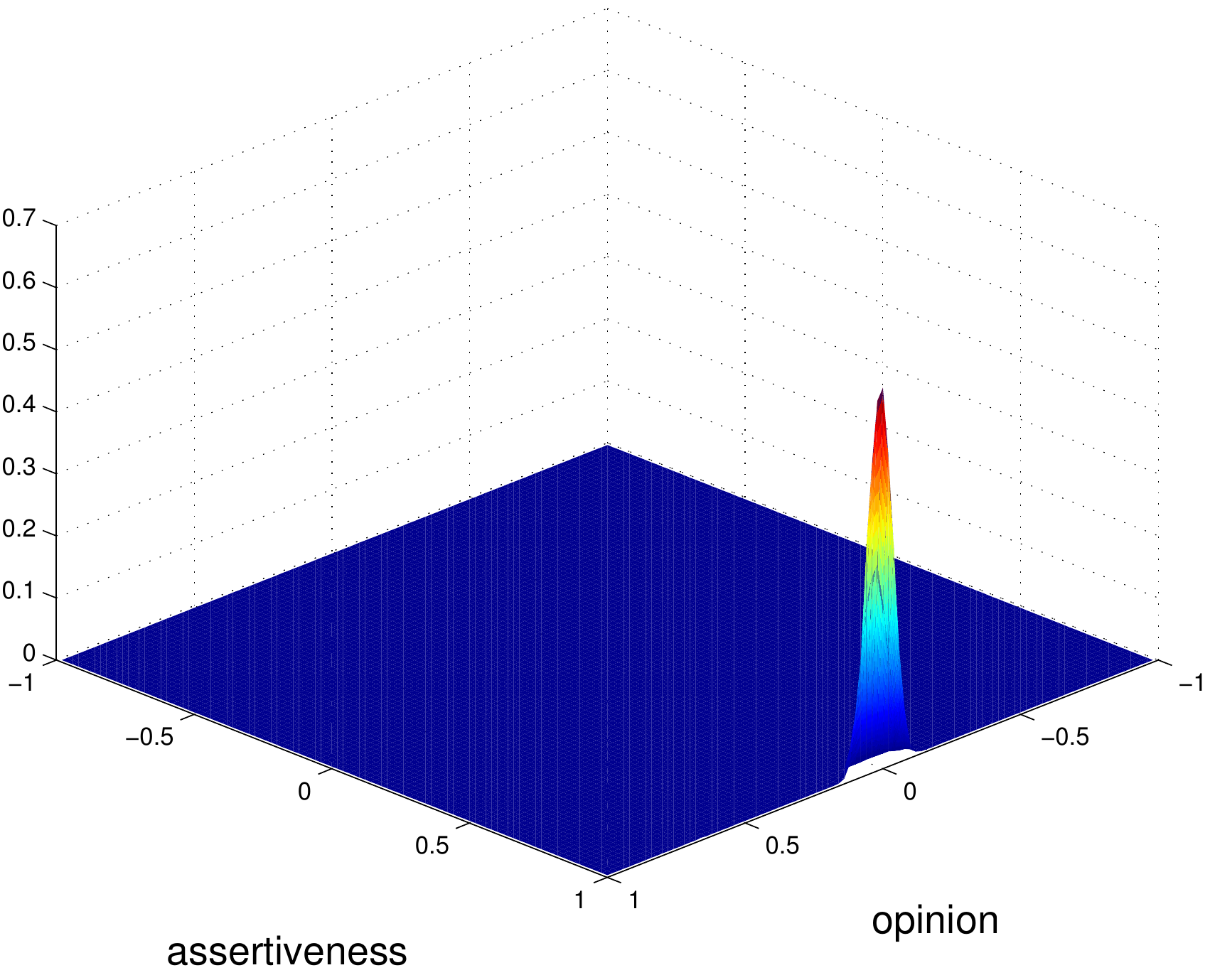}}
\subfigure[Inhomog. $r=0.5$]{\includegraphics[width=0.475\textwidth]{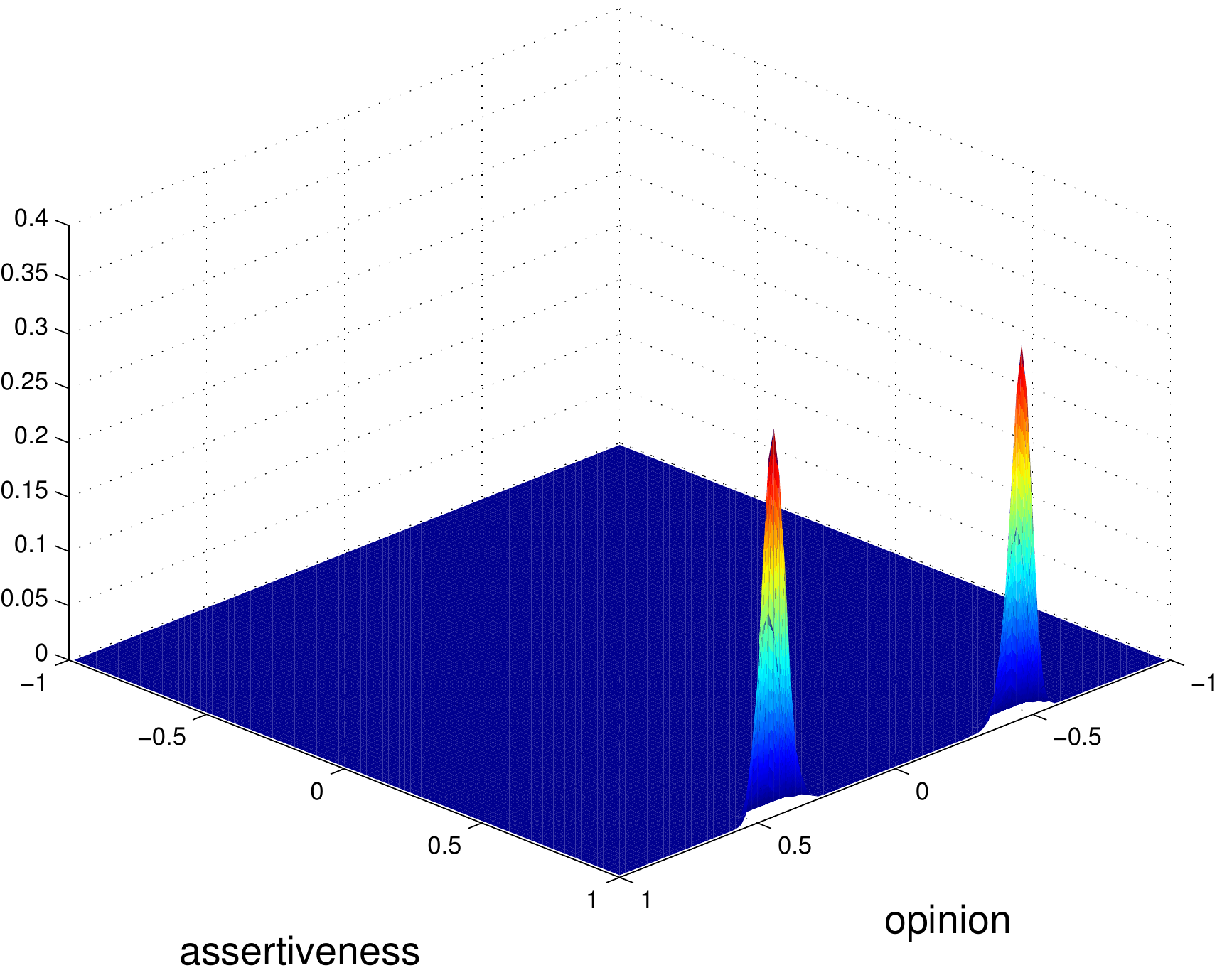}}\\
\subfigure[Multi-dim. $r=1$]{\includegraphics[width=0.475\textwidth]{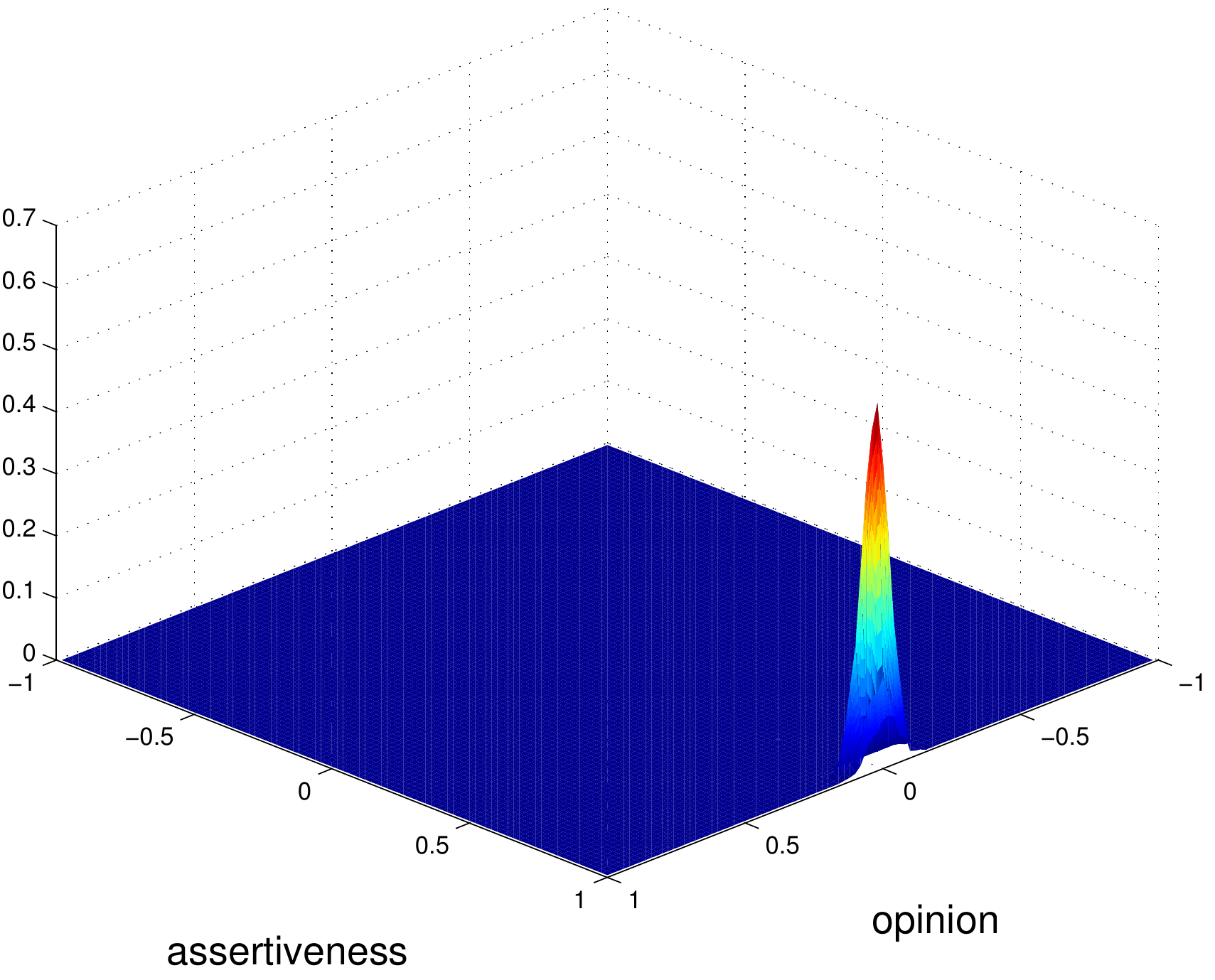}}
\subfigure[Multi-dim $r=0.5$]{\includegraphics[width=0.475\textwidth]{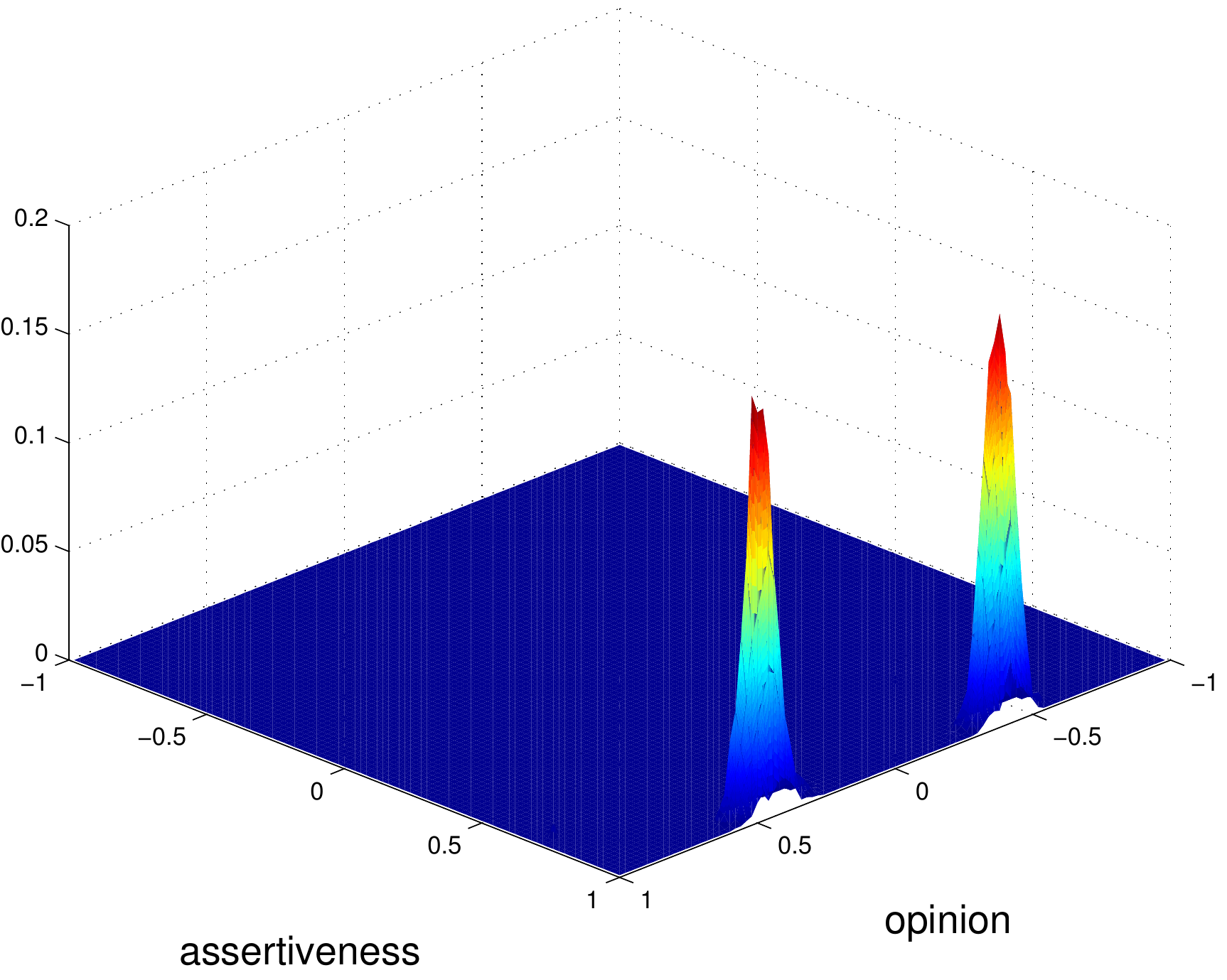}}
\caption{Example 1: Steady state densities of the inhomogeneous and
  multidimensional Boltzmann-type opinion formation model for
  different interaction radii $r$.}\label{f:ex1}.
\end{center}
\end{figure}

First we would like to illustrate the influence of the interaction
radius $r$. We assume 
that the functions in both models which relate to the increase or
decrease of the assertiveness (see \eqref{e:bigsortphi} and \eqref{e:Gtilde})
define a constant increase of the individual assertiveness level, i.e.
\begin{align*}
G(x,w) = 1 \text{ and } \tilde{G}(x,x-y) = (1-x^2)^\alpha\tanh\big(k (x-y)\big),
\end{align*}
with $k=10$. We start each Monte Carlo simulation with an equally distributed number of individuals within $(v,x)  \in (-1,1) \times (-1,1)$. 
We expect the formation of one or several peaks at the highest assertiveness level due to the constant increase caused
by the functions $G$ and $\tilde{G}$. Figure \ref{f:ex1}, which shows the averaged steady state densities for $\delta = \gamma = 0.25$ and two different radii, $r = 0.5$ and $r = 1$. For $r=1$ we observe the formation of a 
single peak located at the highest assertiveness level, $w=1$, in Figure \ref{f:ex1} in the inhomogeneous and multi-dimensional model. In the case of the smaller 
interaction radius, $r=0.5$, two peaks form at the highest assertiveness level. These results are in accordance with numerical simulations of the original Toscani
model \eqref{e:toscani}. 

\subsubsection{Example 2: Different choices of $G$ and $\tilde{G}$}
\begin{figure}[h!]
\begin{center}
\subfigure[Inhomog. $r=1$]{\includegraphics[width=0.475\textwidth]{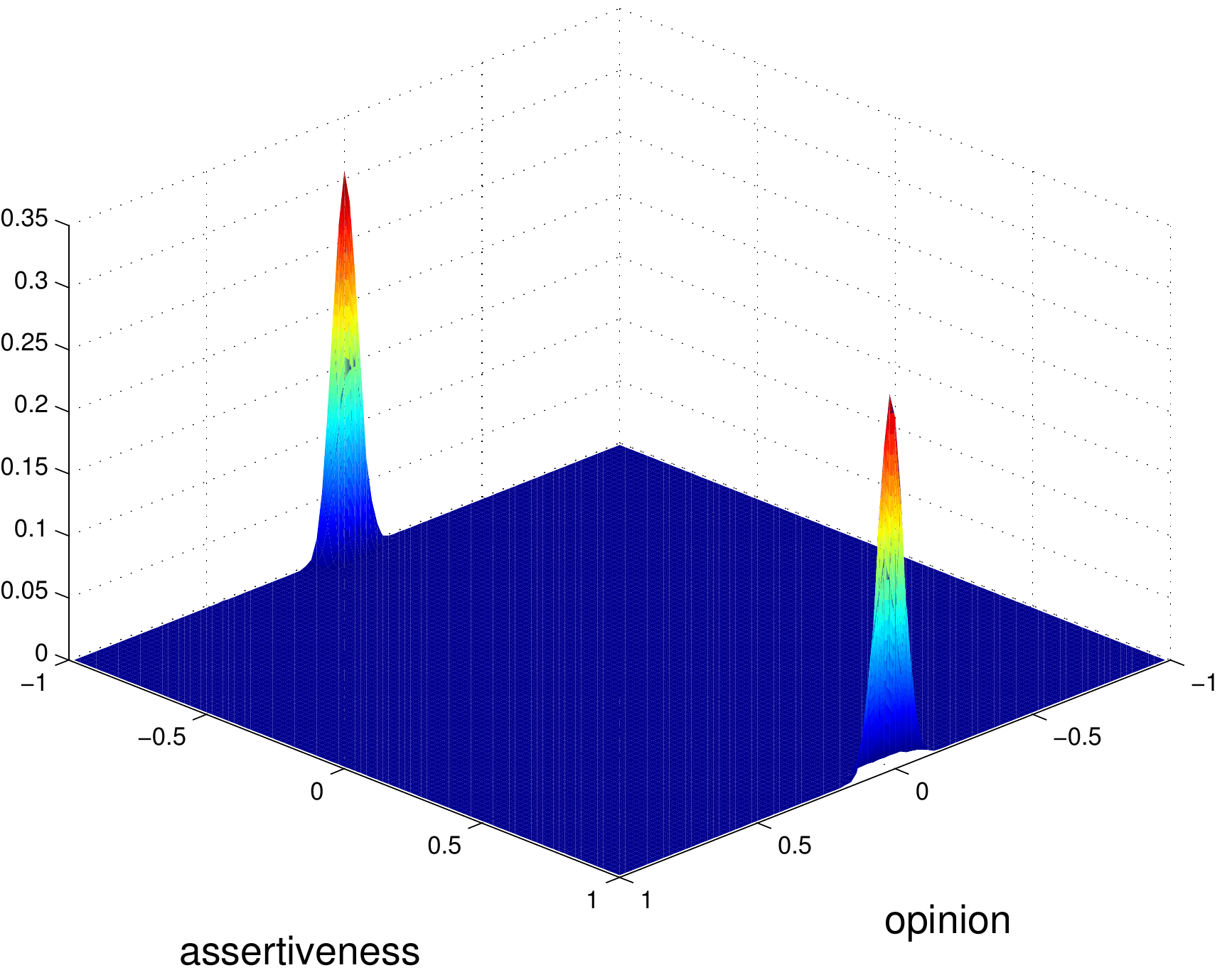}}
\subfigure[Inhomog. $r=0.5$]{\includegraphics[width=0.475\textwidth]{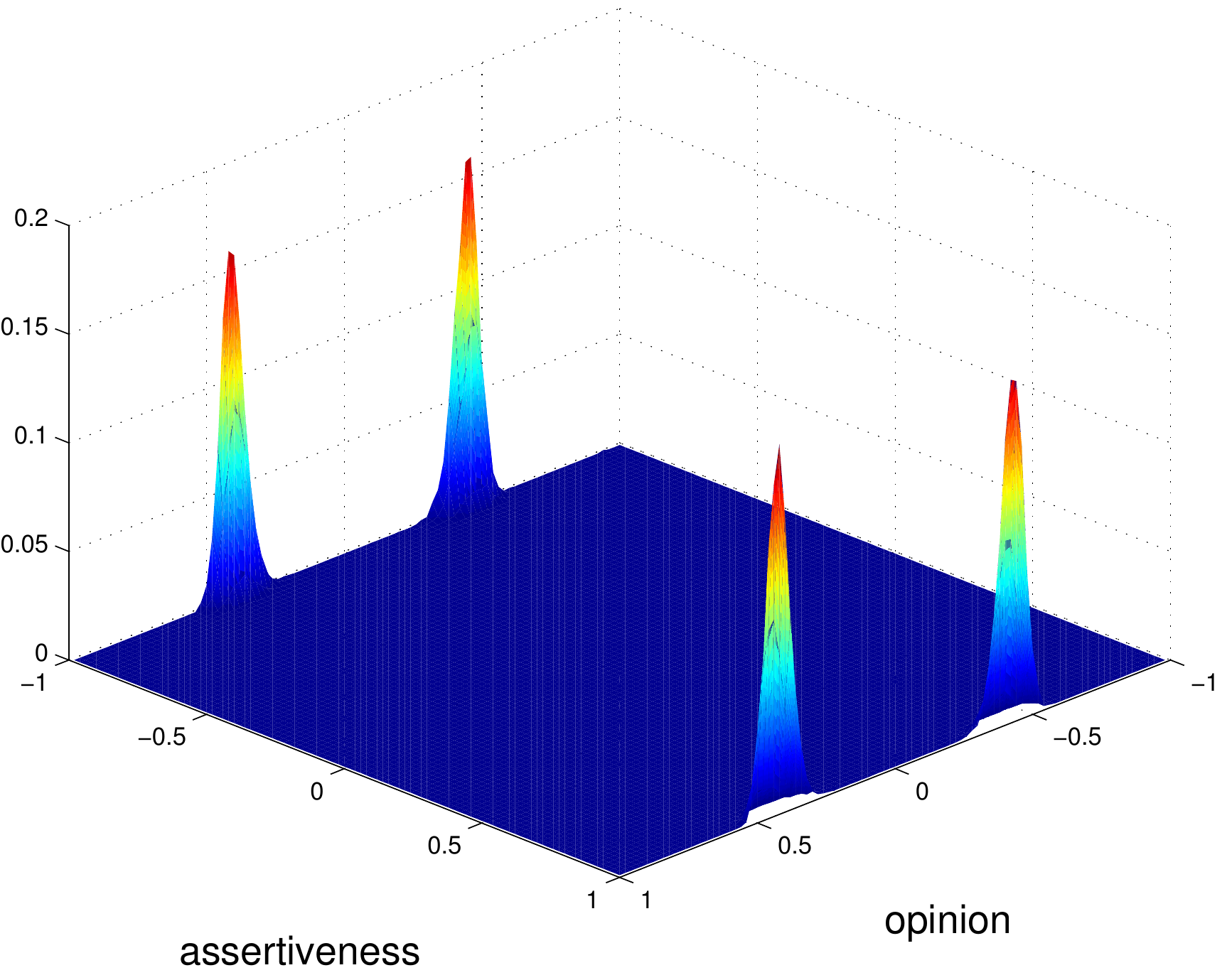}}\\
\subfigure[Multi-dim. $r=1$]{\includegraphics[width=0.475\textwidth]{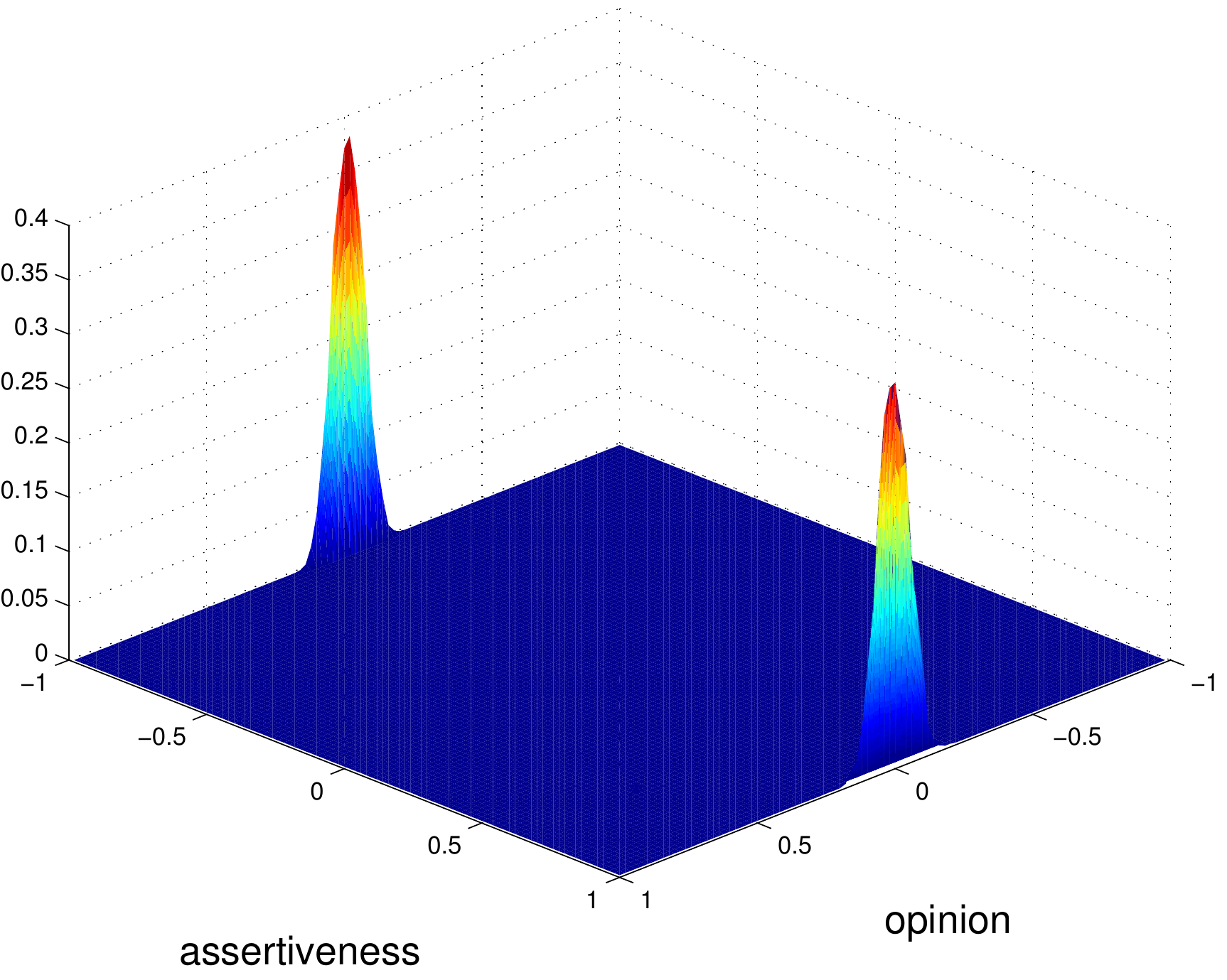}}
\subfigure[Multi-dim $r=0.5$]{\includegraphics[width=0.475\textwidth]{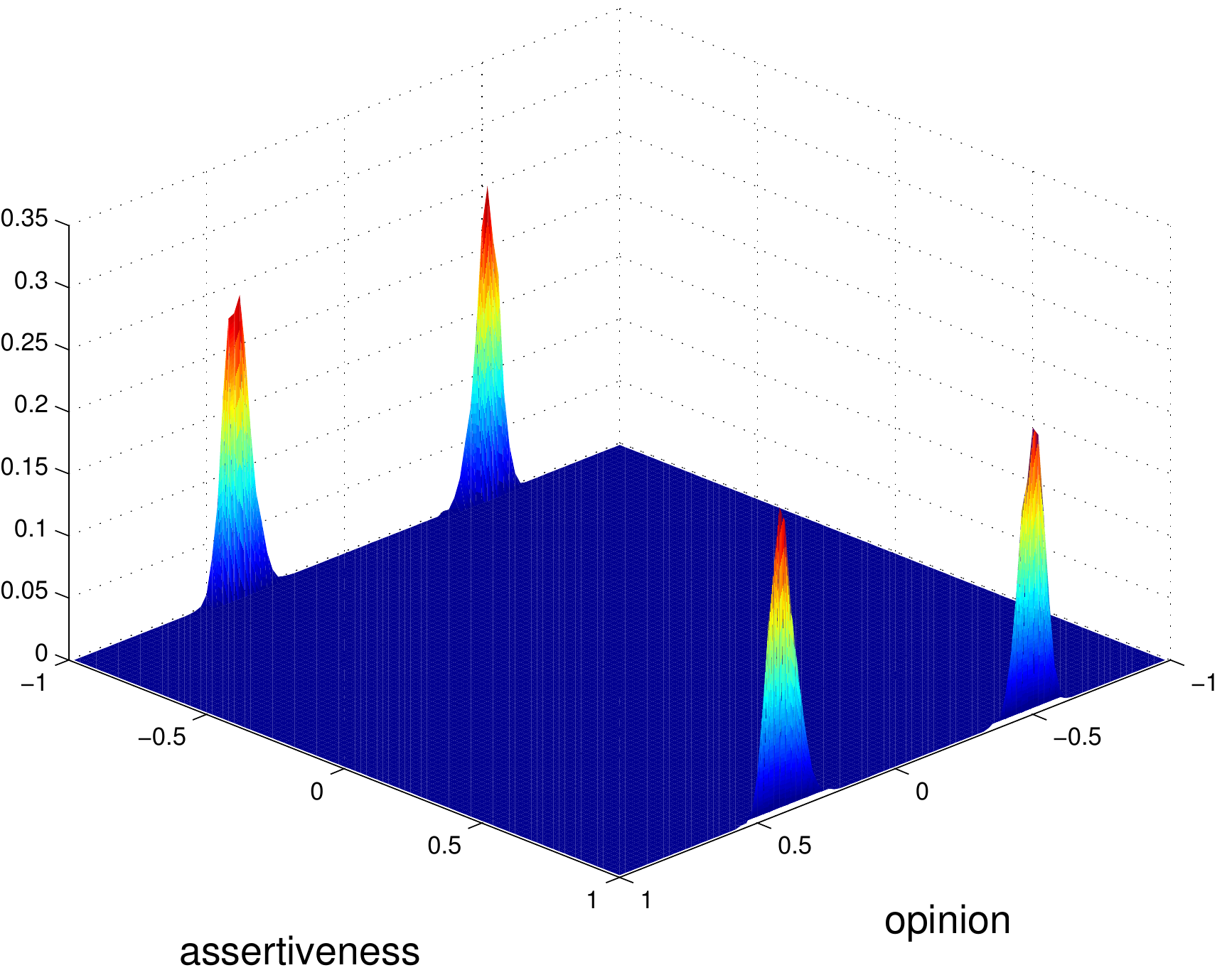}}
\caption{Example 2: Steady state densities of the inhomogeneous and
  multidimensional Boltzmann-type opinion formation model for
  different interaction radii.}\label{f:ex2}.
\end{center}
\end{figure}

Next we focus on the influence of the functions $G$ and $\tilde{G}$,
which model the increase of the assertiveness level in either model (see \eqref{e:bigsortphi} and \eqref{e:Gtilde}).
We choose the same initial distribution as in Example 1 and set
\begin{align*}
G(x,w) = \tanh(k x) \text { and } \tilde{G}(x,x-y) = (1-x^2)^\alpha\lvert x-y \rvert.
\end{align*}
Both functions are based on the following assumption: individuals with
a high assertiveness level gain confidence, while those with a lower
level loose. We expect the formation of peaks close to the highest and lowest assertiveness level. This assumption is confirmed by
our numerical experiments, see Figure \ref{f:ex2}. We observe the formation of peaks close to the maximum and minimum assertiveness
level; the number of peaks again depends on the interaction radius $r$.

\subsubsection{Example 3: Preferred opinions promoting leadership}
Our last example illustrates the rich behaviour of the proposed models. We assume that leadership qualities are directly related
to the opinion of an individual. Individuals with a popular, 'mainstream' opinion, i.e.\ $w = \pm  0.5$ gain confidence, while extreme opinions
are not promoting leadership. To model this we set
\begin{align*}
G(x,w)  = \tilde{G}(x,w,x-y)  =  \frac{1}{\sqrt{2 \pi} \sigma} \left(e^{-\frac{1}{2\sigma^2}(w-0.5)^2} + e^{-\frac{1}{2\sigma^2} (w+0.5)^2}\right)
\end{align*}
with $\sigma^2={1}/{8}$
and 
\begin{align}\label{e:choiceR}
  R(x,x-y) = \Big(\frac{1}{2} - \frac{1}{2} \tanh\big(k(x-y)\big)\Big) e^{-(x+1)^2}
\end{align}
with $k=10$.
The second factor in \eqref{e:choiceR} models the assumption that individuals change their opinion more if they have a low
assertiveness level. 

\begin{figure}[h!]
\begin{center}
\subfigure[Inhomog. ]{\includegraphics[width=0.475\textwidth]{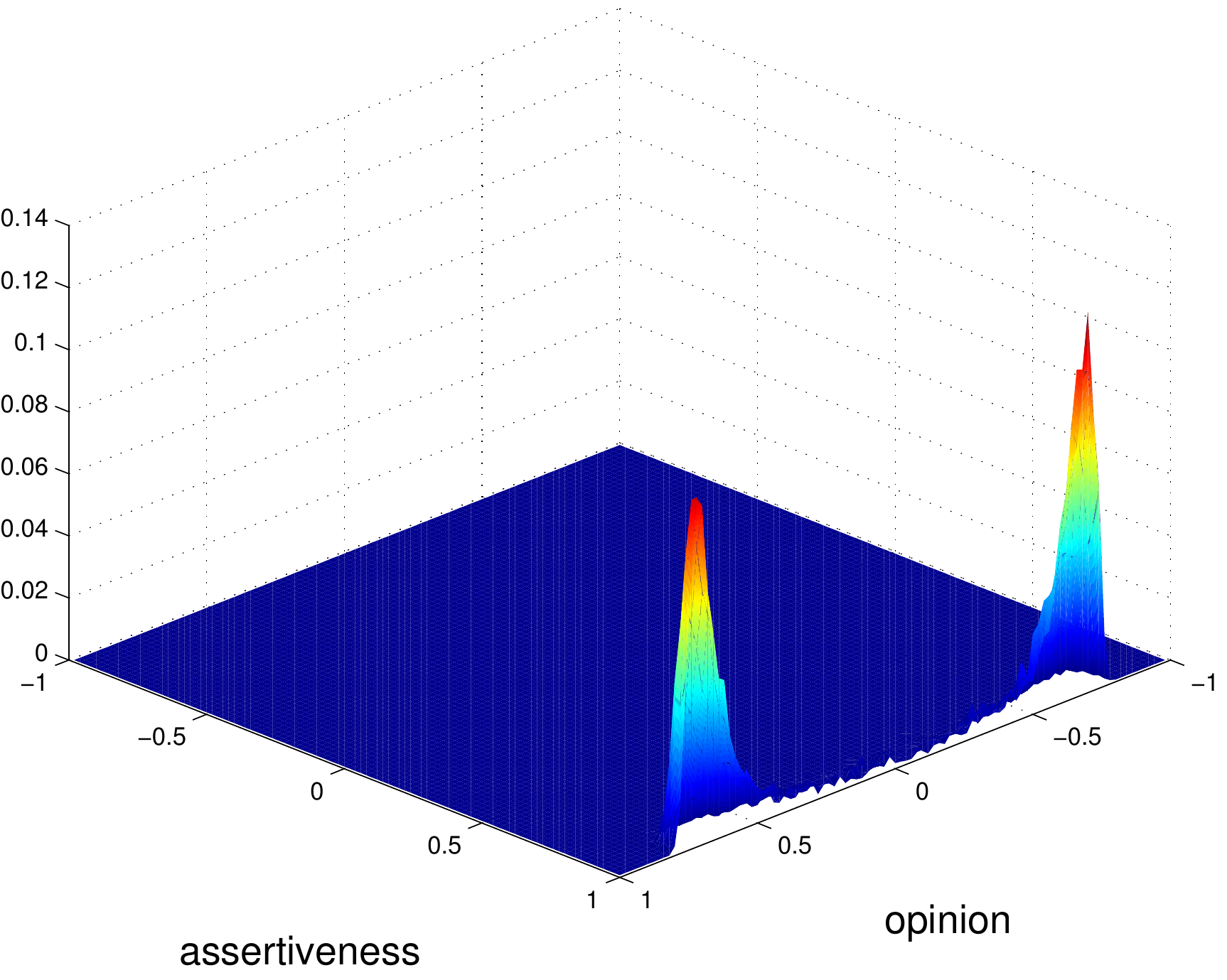}}
\subfigure[Multi-dim.] {\includegraphics[width=0.475\textwidth]{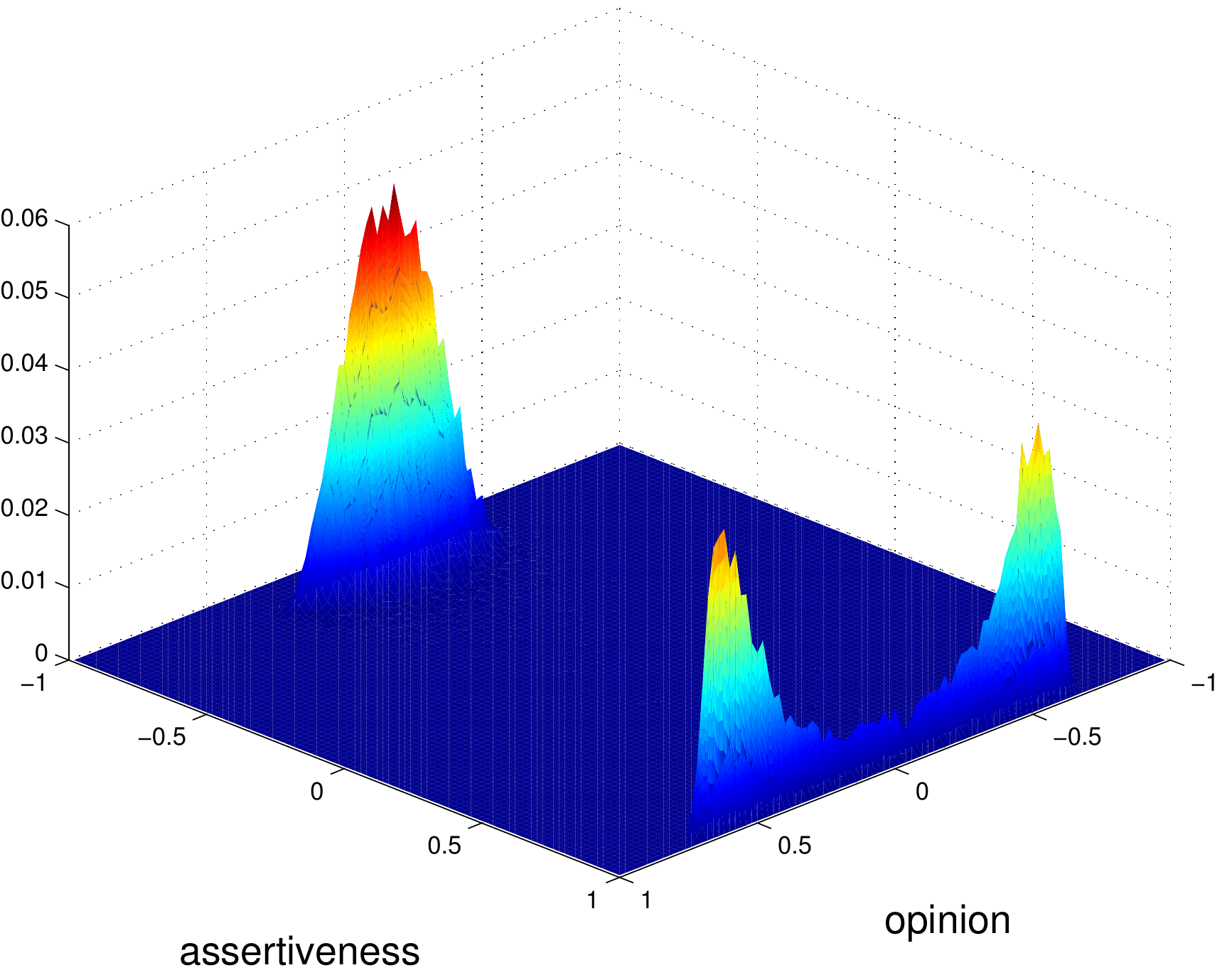}}
\caption{Example 3: Steady state densities of the inhomogeneous and
  multidimensional Boltzmann-type opinion formation model.}\label{f:ex3}.
\end{center}
\end{figure}

At time $t=0$ we equally distribute the individuals within $(w,x) \in [-0.25, 0.25] \times[-0.75,-0.25]$ -- hence we assume that initially no extreme opinions exist and no leaders are present. The interaction radius is set to $r=1$.
 Figure \ref{f:ex3} illustrates the very interesting behaviour in this case. In the multidimensional simulation we observe the
formation of a single peak at a low assertiveness level, while individuals with a higher assertiveness level group around
the `mainstream' opinion $w = \pm 0.5$. In the case of the inhomogeneous model, the potential $\Phi$ initiates an increase of the
assertiveness level for all individuals. Therefore we do not observe the formation of a single peak at a low assertiveness level in this case.

\subsection{Numerical results: `The Big Sort' in Arizona}

\begin{figure}
\begin{center}
\subfigure[1992]{\includegraphics[width=0.2\textwidth]{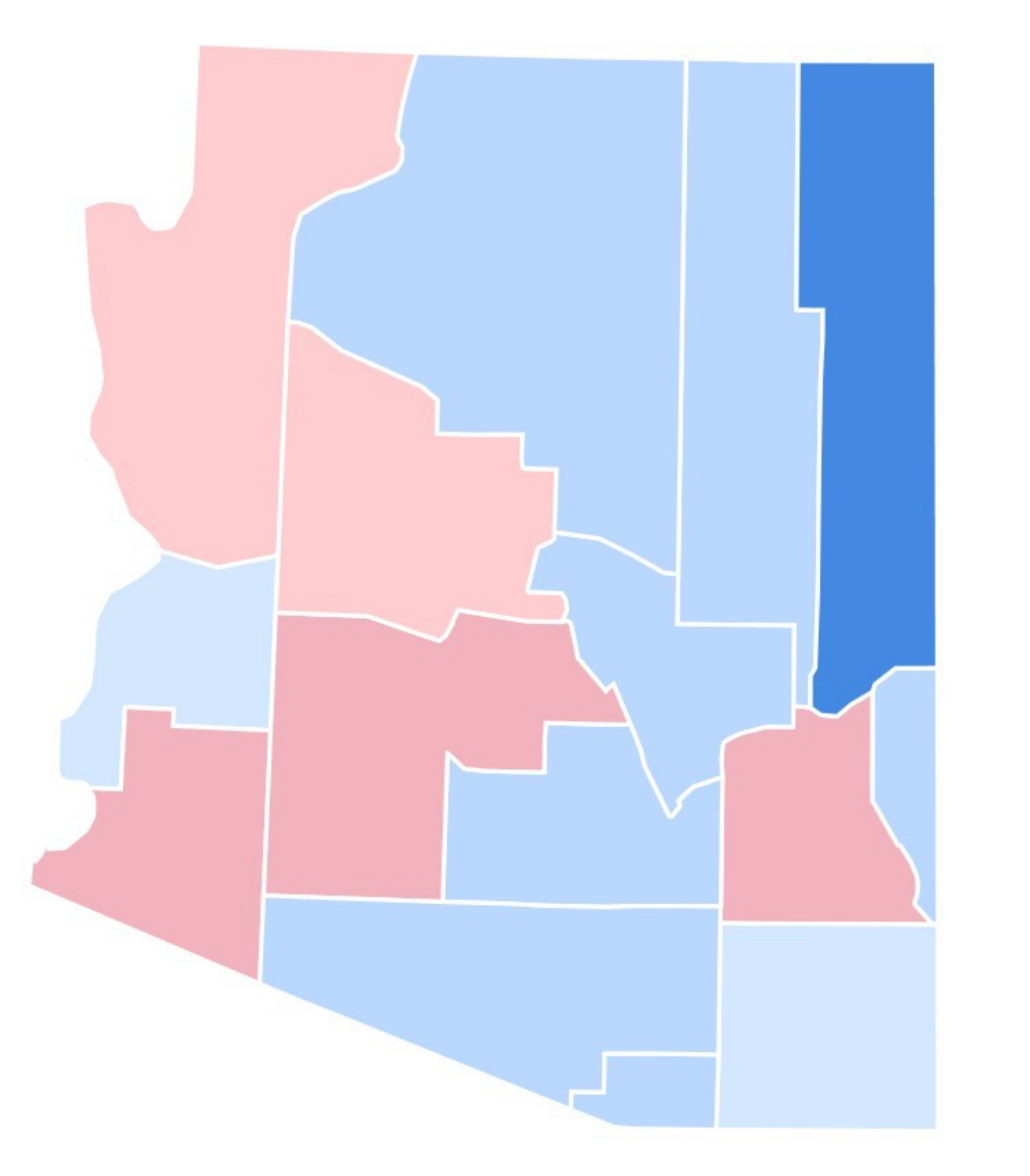}}\hspace*{0.25cm}
\subfigure[1996] {\includegraphics[width=0.2\textwidth]{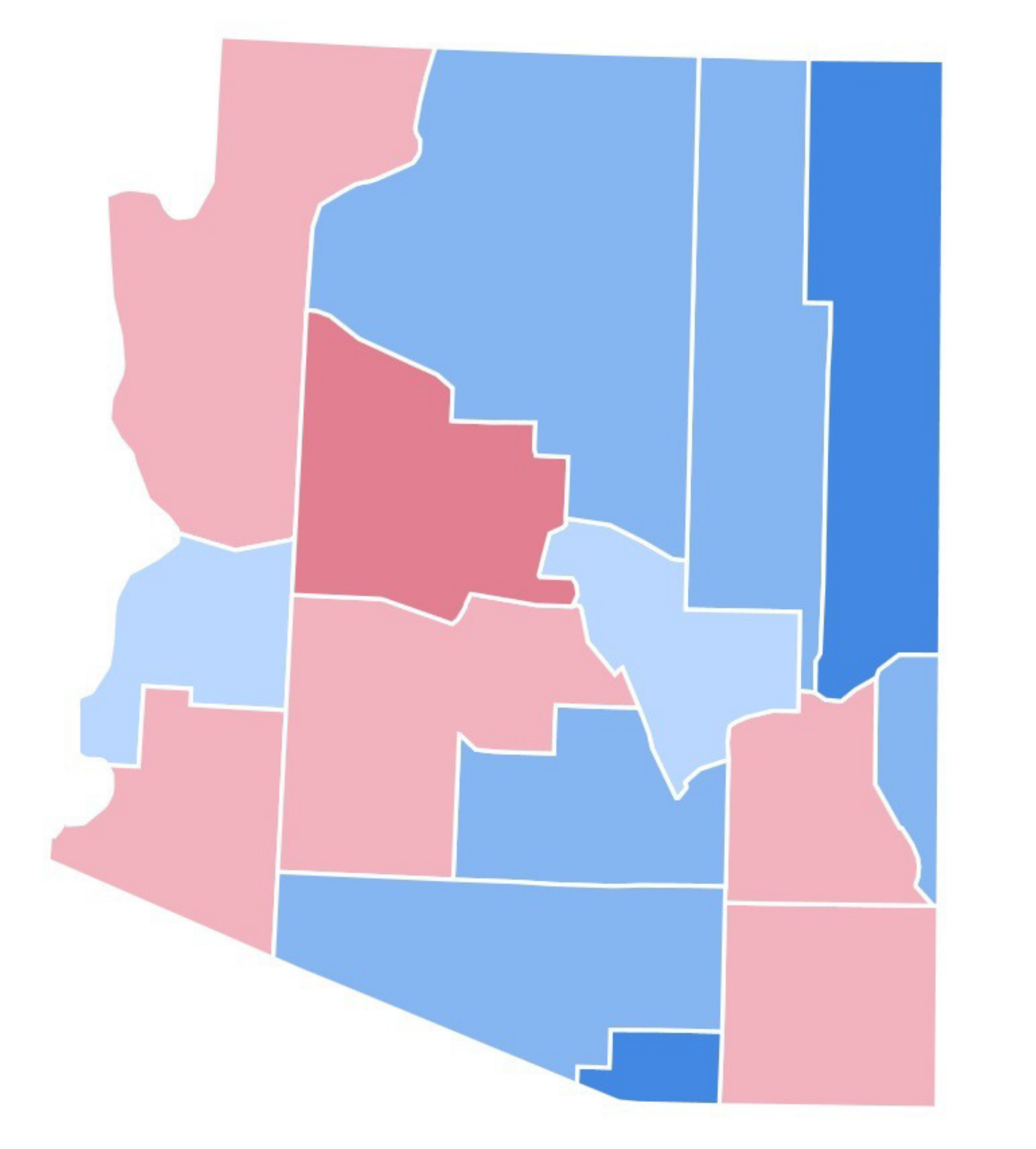}}\hspace*{0.25cm}
\subfigure[2000]{\includegraphics[width=0.225\textwidth]{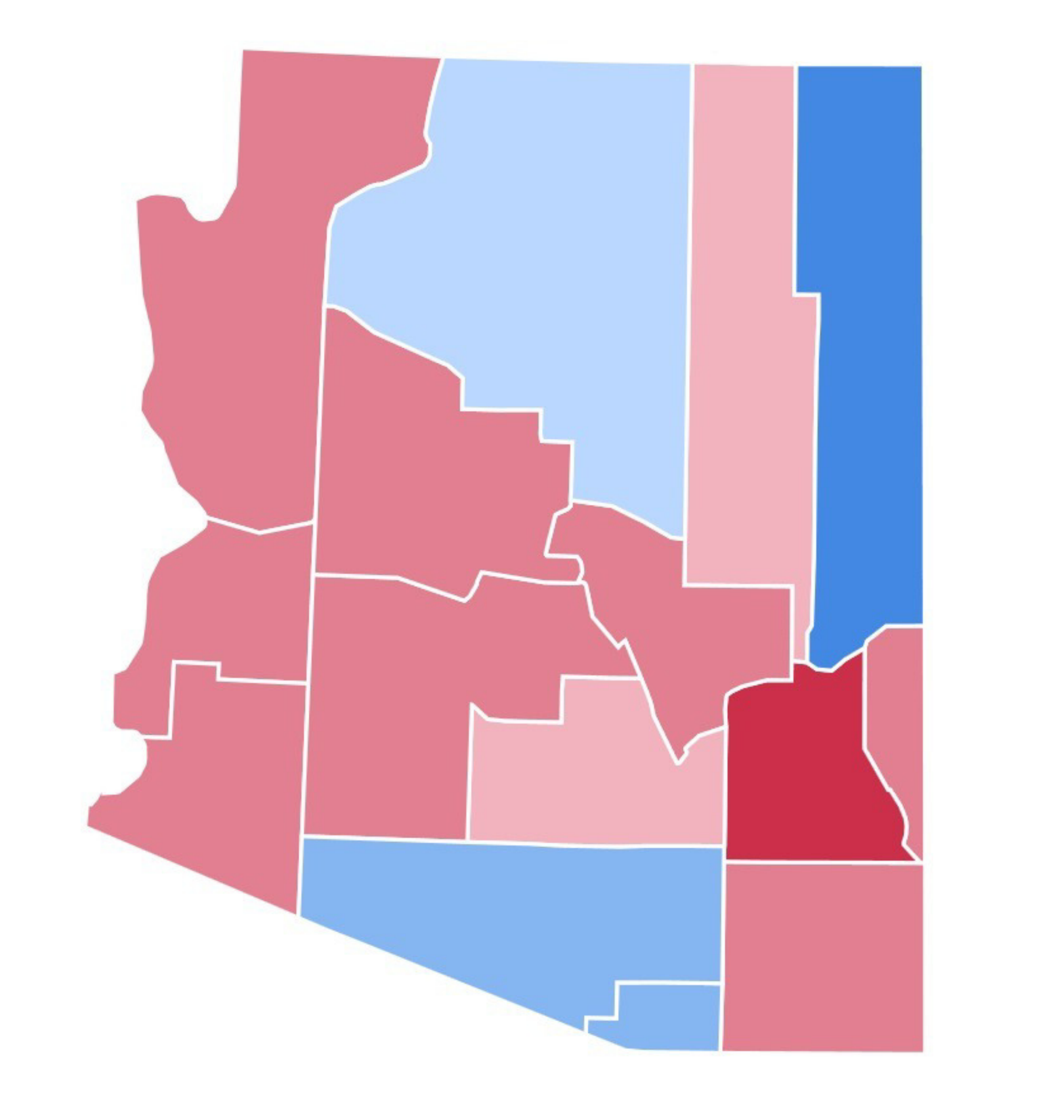}}\hspace*{0.25cm}
\subfigure[2004] {\includegraphics[width=0.225\textwidth]{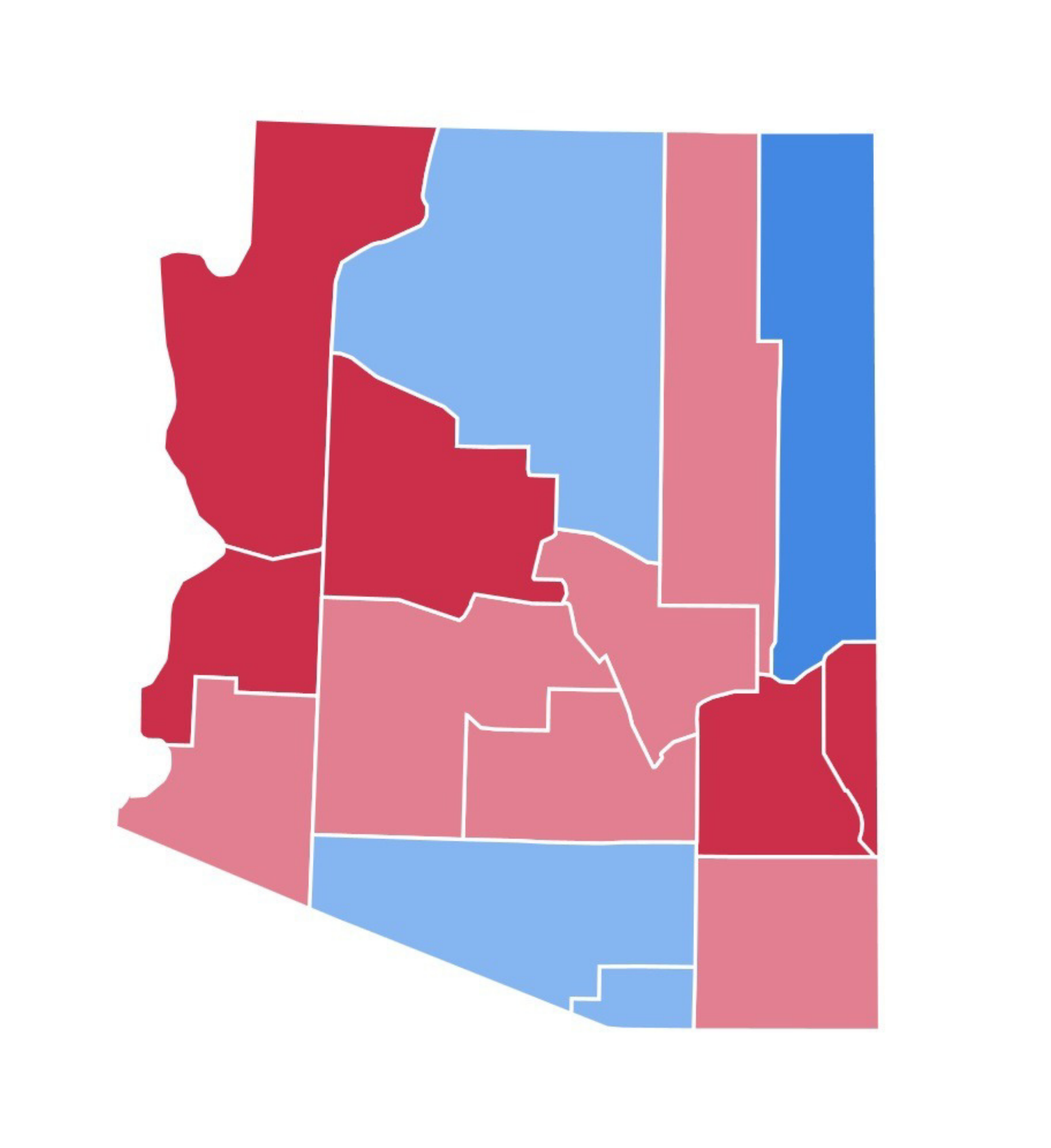}}
\caption{Results of the presidential elections in Arizona \cite{imageswiki}. The colour
  intensities reflect the election outcome in percent, i.e.\ dark blue
  (red) to Democrats (Republicans) $60$--$70\%$, medium blue (red) to
  Democrats (Republicans) $50$--$60 \%$ and light blue (red) to Democrats
 (Republicans) $40$--$50 \%$.}\label{f:elections}.
\end{center}
\end{figure}

In our final example we present numerical simulations of the corresponding Fokker-Planck equation \eqref{e:fokkerplanck}, which illustrate `The Big Sort' by considering the state of Arizona. Arizona is a state in the southwest of the United States with 15 electoral counties. In  recent
years the Republican Party dominated Arizonas politics, see for example the outcome of the presidential elections from the years 1992 to 2004 in
Figure~\ref{f:elections}. The colours red and blue correspond to
Republicans and Democrats, respectively. The colour intensity reflects
the
election outcome in percent, i.e.\ dark blue corresponds to Democrats
$60$--$70\%$, medium blue to Democrats $50$--$60 \%$ and light blue to
Democrats $40$--$50 \%$. Similar colour codes are used for the Republicans. The election results illustrate the clustering trend as the election results per county 
become more and more pronounced over the years.

\begin{figure}
\begin{center}
\subfigure[Computational domain $\Omega$]{ \includegraphics[width=0.225\textwidth]{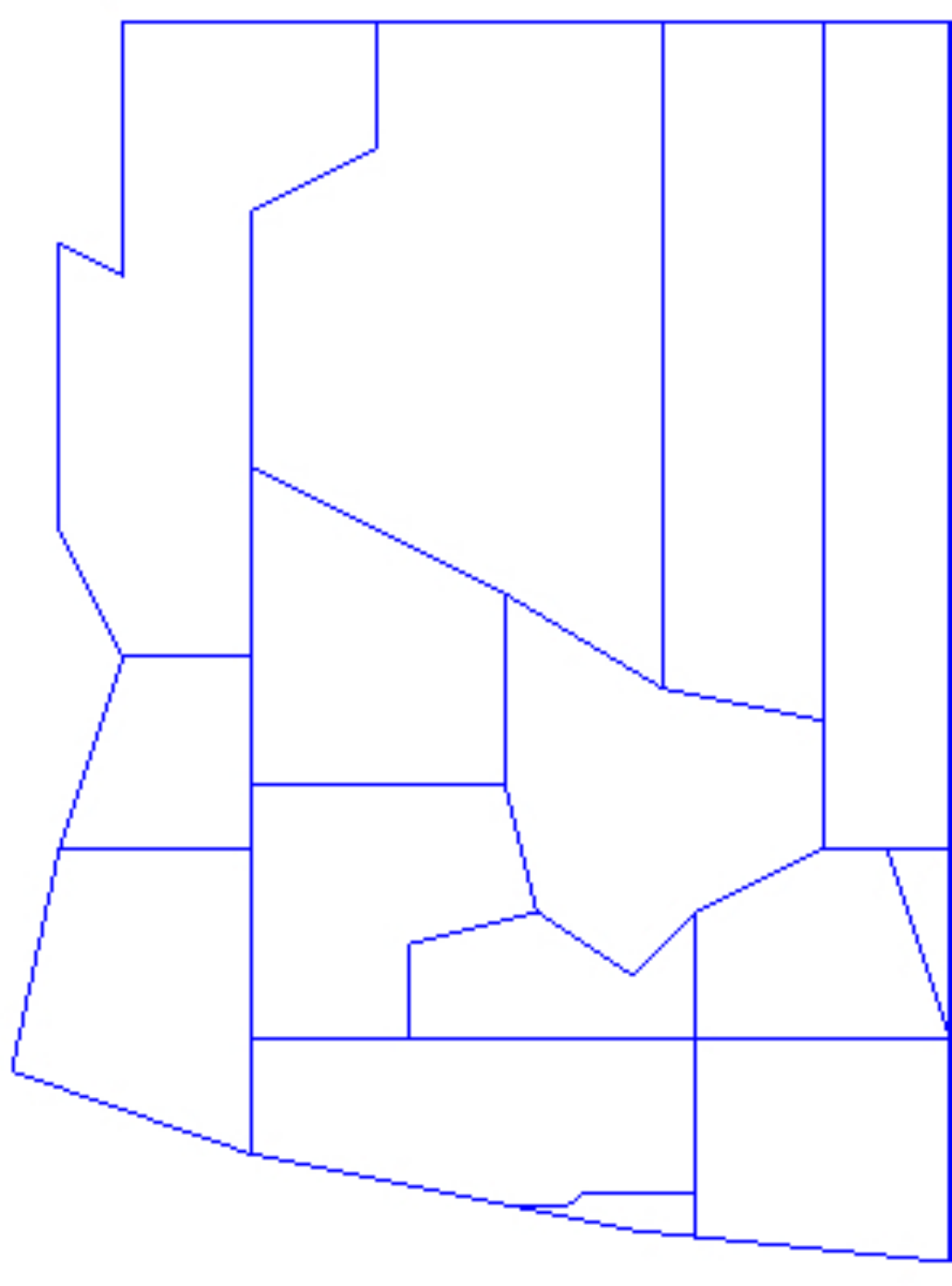}\label{f:domainarizona}}\hspace*{1.5cm}
\subfigure[Potential $C$.]{\includegraphics[width=0.45\textwidth]{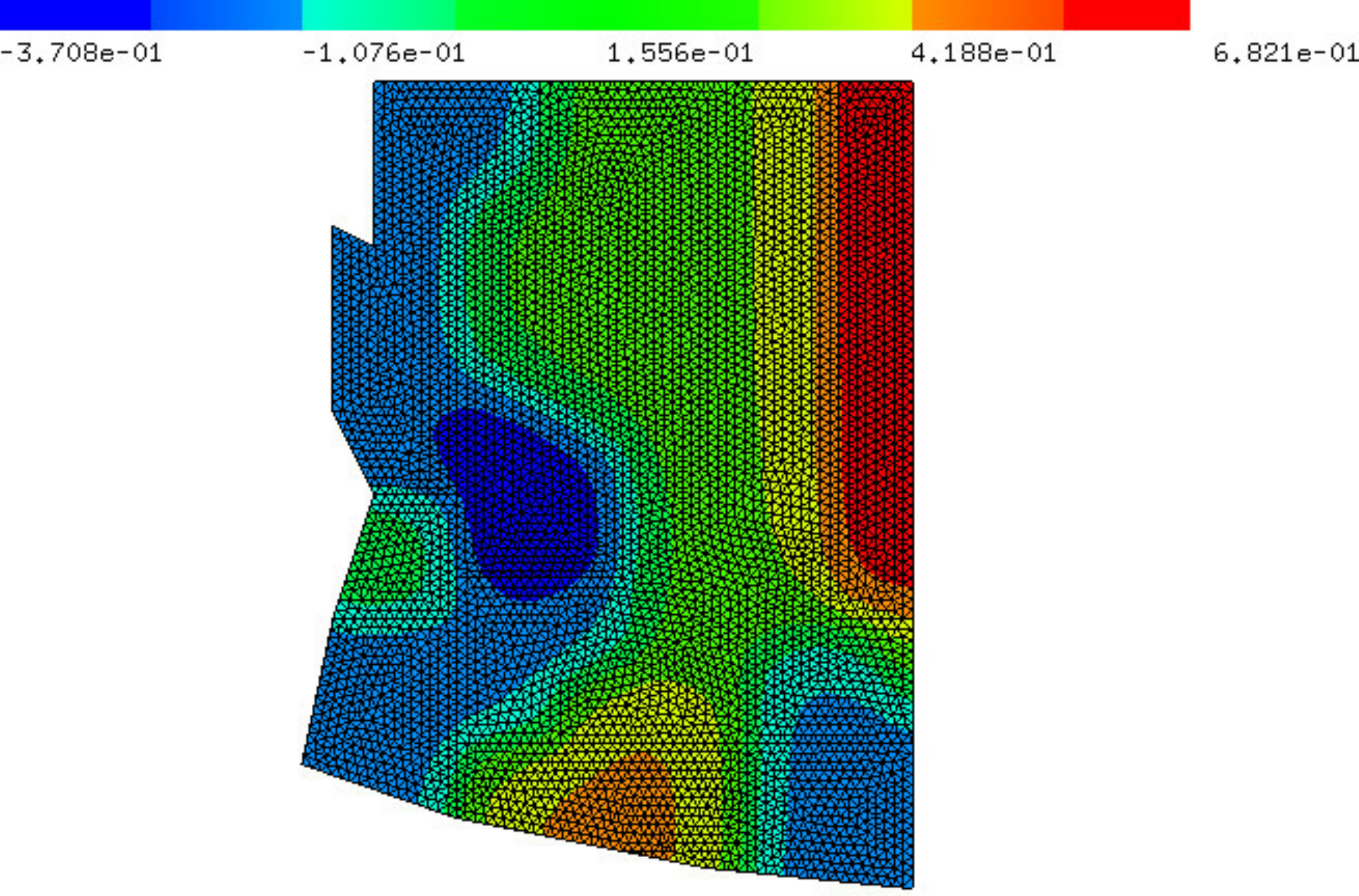}\label{f:pot}}
 \caption{Computational domain $\Omega$ representing the 15 electoral counties of Arizona and potential $C$ corresponding to the election results in 1996.}\label{f:potential}	
 \end{center}
\end{figure}

We solve the Fokker-Planck equation \eqref{e:fokkerplanck} on a
bounded physical domain $\Omega\subset\mathbb{R}^2$ corresponding
to the state Arizona, divided into $15$ electoral counties, see Figure~\ref{f:domainarizona}.
We employ the numerical strategy outlined in Section~\ref{sec:numerics}.\ref{s:simfp}. We discretise the physical domain in $9356$ triangles, the opinion domain $\mathcal{J} = [-1,1]$ into $100$ elements.
The time steps are set to $\Delta t=2\times 10^{-4}$.\\
We choose an initial distribution which is proportional to the election result in 1992,
\begin{align*}
f(x,w,0) = 0.5 + 
\begin{cases}
w(1-w)f_{D,1992}(x) ~~\text{ for } w > 0,\\
w(1+w)f_{R,1992}(x) ~~\text{ for } w < 0,
\end{cases}
\end{align*}
where $f_{D,1992}$ and $f_{R,1992}$ correspond to the distribution of Democrats and Republicans estimated from the election results in $1992$. 
We approximate the distributions by assigning different constants to the respective percentages in the electoral vote, i.e. 
\begin{align}\label{e:elecresults}
 f_{D,R}(x) = 
 \begin{cases}
  0.25 &\text{if the election outcome is } < 40-50\%\\
  0.5 &\text{if the election outcome is between } 50-60 \%\\
  0.75 &\text{ if the election result is } > 60 \%.
 \end{cases}
\end{align}
\noindent The potential $\Phi$, given by \eqref{eq:potential}, attracts individuals to the counties controlled by the party they support. We assume that
the potential $C = C(x)$ is directly related to the electoral results of the year 1996, and satisfies the following PDE:
\begin{align*}
C(x) + \varepsilon \Delta C(x) = f_{1996}(x) ~~\text{ for all } x \in \Omega,\quad
\nabla C(x) \cdot n = 0 ~~ \text{ for all } x \in \partial \Omega.
\end{align*}
The Laplacian is added to smooth the potential $C$, the right hand side corresponds to the election results in the year 1996 (using
the same constants as in \eqref{e:elecresults} with $\pm$ sign corresponding to the Republicans and Democrats respectively). 
Note that we choose Neumann boundary conditions to ensure that individuals stay inside the physical domain. The calculated
potential $C$ is depicted in Figure~\ref{f:pot}. The simulation parameters are set to $\lambda = 0.1$, $\tau = 0.25$ and the interaction radius to $r =5$.

\begin{figure}
\begin{center}
\subfigure[Distribution of democrats]{\includegraphics[width=0.45 \textwidth]{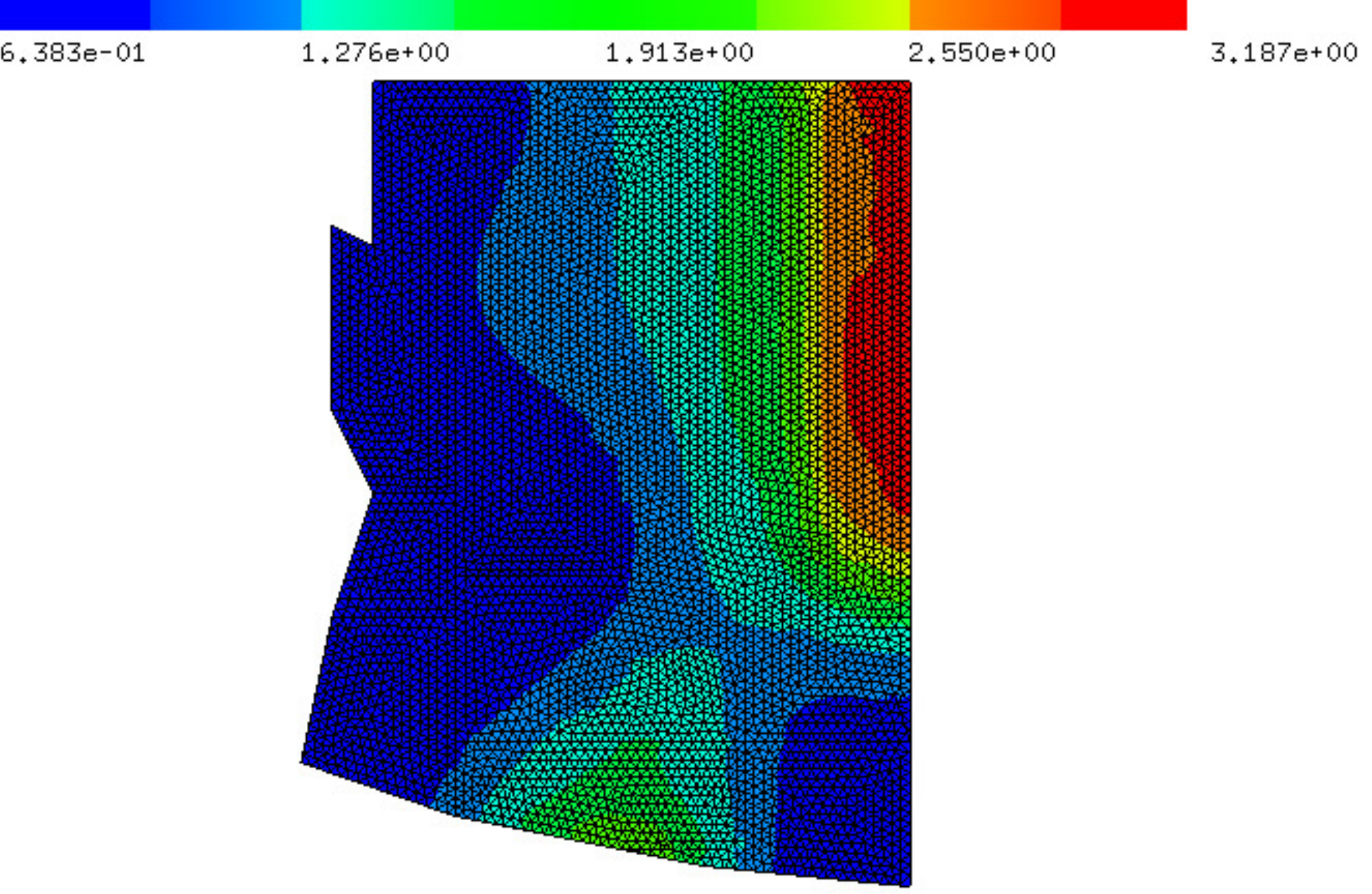}}\hspace*{0.2cm}
\subfigure[Distribution of republicans]{\includegraphics[width=0.45 \textwidth]{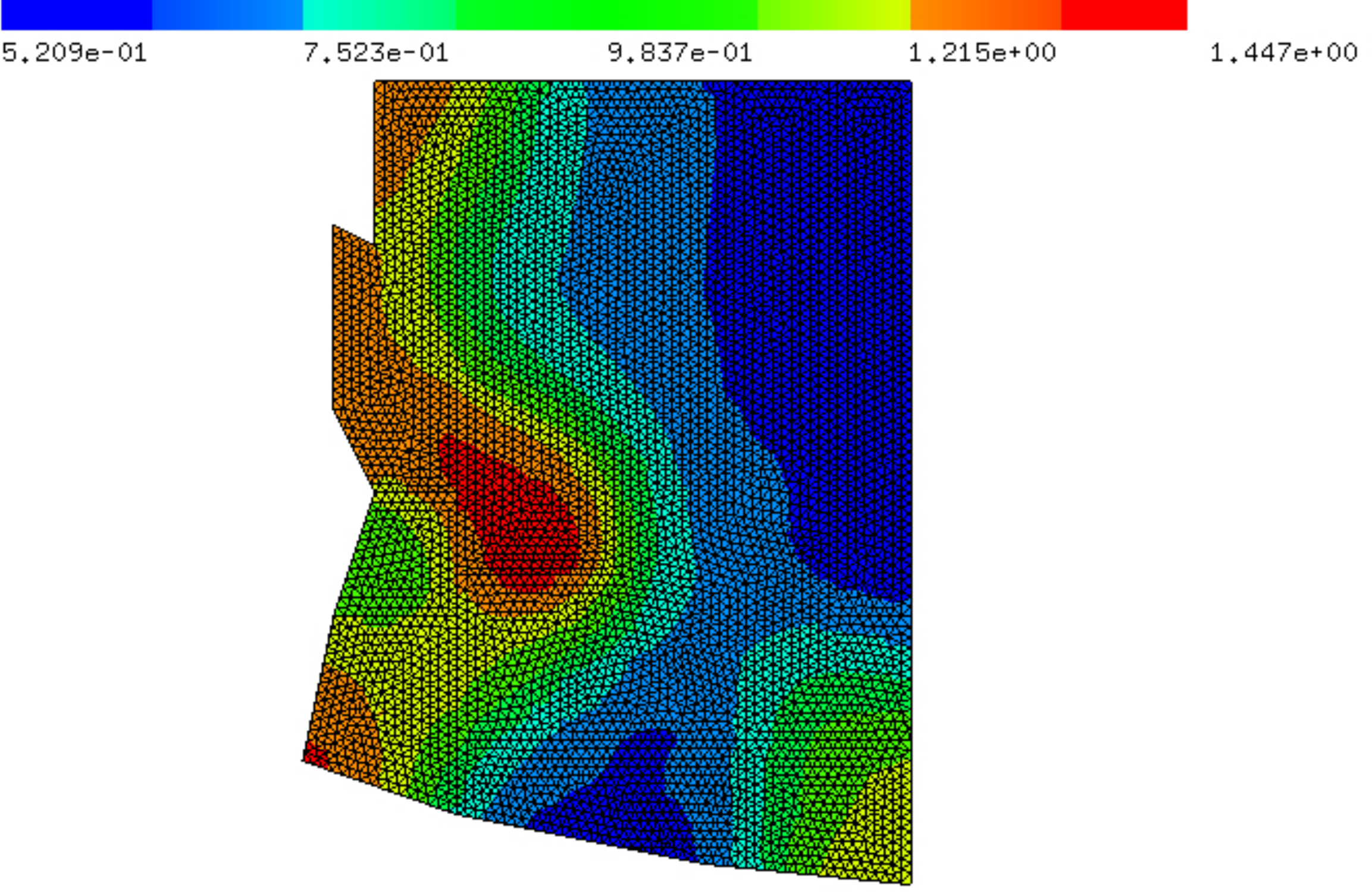}}
\caption{The Big Sort: distribution of Democrats, $f_D(x,t) = \int_{-1}^0 f(x,w,t) \,dw,$ and Republicans, $f_R(x,t) = \int_0^1 f(x,w,t) \,dw$.}\label{f:bigsort}
\end{center}
\end{figure}

Figure \ref{f:bigsort} shows the spatial distribution of the marginals \eqref{e:RB}, i.e.
\begin{align*}
f_D(x,t) = \int_{-1}^0 f(x,w,t)\, dw \quad \text{ and } \quad f_R(x,t) = \int_0^1 f(x,w,t) \,dx,
\end{align*} 
which correspond to Democrats and Republican, respectively, at time
$t=0.5$. 
We observe the formation of two larger clusters of democratic voters in the south and the
Northeast of Arizona. The republicans move towards the Northwest as well as the Southeast. This simulation results reproduce the patterns of the electoral results
from 2004 fairly well, except for the second county from the right in
the Northeast (Navajo). However, given the electoral data from 1992 and 1996 only it is not possible to 
initiate such opinion dynamics. This would require more in-depth information such as the demographic distribution within the counties or the incorporation of several 
sets of electoral results. We leave such extensions for future research.

\section{Conclusion}
\label{sec:concl}

We proposed and examined different inhomogeneous kinetic models for opinion
formation, when the opinion formation process 
depends on an additional independent variable. Examples included
opinion dynamics under the effect of opinion leadership and opinion
dynamics modelling political segregation.
Starting from microscopic opinion consensus dynamics we derived Boltzmann-type
equations for the opinion distribution. In a quasi-invariant opinion
limit they can be approximated by macroscopic Fokker-Planck-type equations. We
presented numerical experiments to illustrate the models' rich
behaviour. Using presidential election results in the state of Arizona,
we showed an example modelling the process of political segregation in
the `The Big Sort', the process of clustering of individuals who
share similar political opinions. 

\section*{Acknowledgements}
MTW acknowledges financial support from the Austrian Academy of Sciences (\"OAW) via
the New Frontiers Grant NFG0001.\\
The authors are grateful to Prof.\ Richard Tsai (UT Austin) for
suggesting `The Big Sort' as an example for an inhomogeneous
opinion formation process.

\end{document}